\documentclass[aps, pre, 10pt, twocolumn]{revtex4-1}
\usepackage{amsmath, amssymb, bm, graphicx}
\usepackage{mathrsfs}
\usepackage[dvipsnames]{xcolor}
\usepackage{hyperref}
\usepackage[normalem]{ulem}
\hypersetup{colorlinks, citecolor = blue, filecolor = blue, linkcolor = blue, urlcolor = blue}

\begin{document}

\title{Reversals in infinite-Prandtl-number Rayleigh-B\'{e}nard convection}

\author{Ambrish Pandey$^1$}
\email{ambrish.pandey@tu-ilmenau.de}
\author{Mahendra K. Verma$^2$}
\email{mkv@iitk.ac.in}
\author{Mustansir Barma$^3$}
\email{barma@tifrh.res.in}
\affiliation{$^1$Institut f\"ur Thermo- und Fluiddynamik, Technische Universit\"at Ilmenau, Postfach 100565, D-98684 Ilmenau, Germany \\ $^2$Department of Physics, Indian Institute of Technology Kanpur, Kanpur 208016, India \\$^3$TIFR Centre for Interdisciplinary Sciences, Tata Institute of Fundamental Research, Gopanpally, Hyderabad 500107, India}

\date{\today}

\begin{abstract}
Using direct numerical simulations, we study the statistical properties of reversals in two-dimensional Rayleigh-B\'enard convection for infinite Prandtl number. We find that the large-scale circulation reverses irregularly, with the waiting time between two  consecutive genuine reversals exhibiting a Poisson distribution on long time scales, while the interval between successive crossings on short time scales shows a power law distribution. We observe that the vertical velocities near the sidewall and at the center show different statistical properties. The velocity near the sidewall shows a longer autocorrelation and $1/f^2$ power spectrum for a wide range of frequencies, compared to shorter autocorrelation and a narrower scaling range for the velocity at the center. The probability distribution of the velocity near the sidewall is bimodal, indicating a reversing velocity field. We also find that the dominant Fourier modes capture the dynamics at the sidewall and at the center very well. Moreover, we show a signature of weak intermittency in the fluctuations of velocity near the sidewall by computing temporal structure functions.
\end{abstract}

\maketitle

\section{Introduction}

The phenomenon of field reversals occurs commonly in nature, e.g., the polarity of geomagnetic field reverses over a period of several thousand years~\cite{Glatzmaier:NATURE1995}, whereas the large-scale magnetic field in the Sun reverses approximately every 11 years~\cite{Hanasoge:ARFM2016}. Flow reversals are also observed in Rayleigh-B\'{e}nard convection (RBC),   an idealized model of thermal convection, where the large-scale circulation (LSC) switches its direction at irregular intervals~\cite{Sreenivasan:PRE2002, Xi:PRE2006, Brown:JFM2009, Mishra:JFM2011, Ni:JFM2015, Breuer:EPL2009, Sugiyama:PRL2010, Petschel:PRE2011, Chandra:PRE2011, Chandra:PRL2013, Verma:POF2015, Podvin:JFM2015, Yanagisawa:PRE2011}. In RBC a fluid placed in a cylindrical or rectangular container is heated from below and cooled from above~\cite{Ahlers:RMP2009, Chilla:EPJE2012, Verma:NJP2017}. Properties of a convective flow depend primarily on the Rayleigh number $\mathrm{Ra}$ and the Prandtl number $\mathrm{Pr}$. The Rayleigh number signifies the relative strength of the thermal driving force compared to the dissipative forces, and the Prandtl number is the ratio of the kinematic viscosity and the thermal diffusivity of the fluid. Another important governing parameter is the aspect ratio  $\Gamma$, which is the ratio of the width to the height of the RBC cell. In this paper, we study flow reversals in a two-dimensional (2D) square box for $\mathrm{Pr} = \infty$. Infinite Prandtl number RBC can be utilized to model mantle convection in the Earth, where the Prandtl number is extremely large ($\mathrm{Pr} \approx 10^{24}$)~\cite{Schubert:book2001}. 

For RBC in a cylinder with $\Gamma \approx 1$, azimuthal orientation of LSC jitters continuously~\cite{Brown:JFM2006, Xi:PRE2006, Mishra:JFM2011}, and the probability distribution of this angular change decays as a powerlaw~\citep{Brown:PRL2005}. An angular change of approximately $180 \deg$ can be perceived as a reversal of LSC. A reversal can also occur following a cessation event, where the strength of LSC ceases momentarily, and it reappears with a different azimuthal orientation~\cite{Brown:PRL2005, Xi:PRE2006, Mishra:JFM2011}. \citet{Mishra:JFM2011} and \citet{Xi:JFM2016} investigated the reversals of LSC  in a cylindrical geometry, and observed that during a cessation-led reversal, the strength of secondary flow modes increase, whereas that of the primary mode decreases. In 2D RBC, reversals are constrained to occur through a cessation event, and the dominance of secondary modes during a cessation have also been reported~\cite{Breuer:EPL2009, Sugiyama:PRL2010, Chandra:PRE2011, Petschel:PRE2011, Chandra:PRL2013, Verma:POF2015, Podvin:JFM2015}. Moreover, it has been observed that the first few most energetic flow modes capture the flow pattern and dynamics of LSC reversals very well~\cite{Chandra:PRE2011, Petschel:PRE2011, Chandra:PRL2013, Verma:POF2015, Podvin:JFM2015}. In this paper, we gathered very long time statistics of the temporal evolution of vertical velocities at various locations in our simulation domain, and observe that their evolution and statistical properties can be described very well by the most energetic Fourier modes of the flow. 

\citet{Niemela:JFM2001} and  \citet{Sreenivasan:PRE2002} studied  convective reversals for a wide range of $\mathrm{Ra}$ in a $\Gamma = 1$ cylindrical cell filled with cryogenic helium gas ($\mathrm{Pr} \approx 0.7$). They observed that the LSC prefers one direction over the other for $\mathrm{Ra} \lessapprox 10^{11}$, but the probability of being in either direction becomes approximately equal for larger $\mathrm{Ra}$. They found that the waiting time between two consecutive reversals is exponentially distributed for longer waiting times. However, shorter waiting times were observed to be distributed as a powerlaw~\citep{Sreenivasan:PRE2002}. \citet{Brown:JFM2006} however concluded that the powerlaw distribution occurs due to ``crossings" of LSC, which are the reorientation events with angular change of approximately $90 \deg$. They observed that the waiting times between consecutive reversals follow a Poisson distribution for entire range of waiting times~\citep{Brown:JFM2006}, which was also endorsed by the findings of~\citet{Xi:PRE2006} and \citet{Xi:PRE2007, Xi:PRE2008}. In our numerical simulations for infinite Prandtl number, we also observe that the waiting times between successive reversals exhibit exponential distribution, whereas the crossings (to be defined later for the present case) are distributed as a powerlaw. Moreover, the mean waiting time between two consecutive reversals have also been observed to depend on the Rayleigh number~\cite{Niemela:JFM2001, Sreenivasan:PRE2002, Sugiyama:PRL2010, Huang:JFM2016}, the aspect ratio $\Gamma$~\cite{Ni:JFM2015}, and the thermal boundary condition at the bottom plate~\cite{Huang:PRL2015}. 

Another interesting facet of the present investigation is the statistical properties of fluctuations in the time evolution of velocity field recorded at various probes, which we utilize to investigate reversals. For homogeneous and isotropic turbulence, Kolmogorov~\cite{Kolmogorov:DANS1941a} deduced that the third order structure function is proportional to distance between two points in the inertial range. The higher order structure functions are however more complex. This is known as anomalous scaling~\cite{Sreenivasan:ARFM1997, Lohse:ARFM2010}, which arises due to intermittency of viscous dissipation rate. Temporal structure functions have also been utilized to study the anomalous scaling~\cite{Skrbek:PRE2002, Ching:PRE2000a, He:JFM2014}. \citet{Skrbek:PRE2002} computed the temporal structure functions of temperature field recorded near the sidewall of their cylindrical RBC cell filled with cryogenic helium gas at $\mathrm{Ra} = 1.5 \times 10^{11}$, and found the signature of intermittency. \citet{Ching:PRE2003} studied the temporal structure functions of the velocity field at the center of a cylindrical RBC cell filled with water at $\mathrm{Ra} = 3.7 \times 10^9$, and corroborated the anomalous scaling. Moreover, they observed that the velocity structure functions satisfy the She-Leveque scaling~\citep{She:PRL1994}. 

In this paper, we study the properties of infinite-$\mathrm{Pr}$ reversals using the time series of vertical velocity at probes located near a sidewall and at the center of our 2D square box. We also study the evolution of dominant Fourier modes, and find that all the odd-odd modes (the modes whose both indices are odd) are statistically similar to the  vertical velocity at the sidewall probe, as they switch their signs after a flow reversal, their probability distributions are bimodal, and their power spectra exhibit $1/f^{\alpha}$ scaling for a wide range of frequencies. Additionally, by computing the temporal structure functions, we find a signature of intermittency in the fluctuations of the vertical velocity near the sidewall and the most energetic Fourier mode.

The remainder of the paper is organized as follows. In Sec.~\ref{sec:eqns}, we describe the governing equations and numerical method. In Sec.~\ref{subsec:rev_stat}, the statistics of waiting times between consecutive reversals will be discussed. The statistical properties of the time series of the vertical velocities and of the dominant Fourier modes will be presented in Sec.~\ref{subsec:modes}, and intermittency in their fluctuations using structure functions will be examined in Sec.~\ref{subsec:str_fns}. We summarize our main results in Sec.~\ref{sec:conclusion}.

%%%%%%%%%%%%%%%%%%%%%%%%%%%%%%%%%%%%%%%%%%%%%%%%%%%%%%%%%%%%%%%%%%%%%%%%%%%%%%%%%%%%%%%%
%%%%%%%%%%%%%%%%%%%%%%%%%%%%%%%%%%%%%%%%%%%%%%%%%%%%%%%%%%%%%%%%%%%%%%%%%%%%%%%%%%%%%%%%

\section{Governing equations and numerical method} \label{sec:eqns}
Conservation of momentum, energy, and mass lead to equations which govern the dynamics of RBC. For very large Prandtl number~\cite{Pandey:PRE2014, Verma:POF2015}, these equations under the Oberbeck-Boussinesq approximation~\cite{Chandrasekhar:Book, Verma:NJP2017} are
\begin{eqnarray}
\frac{1}{\mathrm{Pr}} \left[ \frac{\partial {\bf u}}{\partial t} + {\bf u} \cdot \nabla {\bf u} \right] & = & -\nabla \sigma + \theta \hat{z} + \frac{1}{\sqrt{\mathrm{Ra}}} \nabla^2 {\bf u}, \label{eq:u} \\
\frac{\partial \theta}{\partial t} + {\bf u} \cdot \nabla \theta & = &  u_z + \frac{1}{\sqrt{\mathrm{Ra}}} \nabla^2 \theta, \label{eq:T} \\
\nabla \cdot {\bf u} & = & 0 \label{eq:m},
\end{eqnarray}
where ${\bf u} \, ( = u_x \hat{x} + u_z \hat{z})$ is the velocity field, and $\theta$ and $\sigma$ are the fluctuations in temperature and pressure from the conduction state. Here $\mathrm{Ra} = \alpha g \Delta d^3/(\nu \kappa)$ and $\mathrm{Pr} = \nu/\kappa$, with $\Delta$ being the temperature difference between the top and bottom plates separated by the distance $d$, $g$ is the acceleration due to gravity, and $\alpha, \nu$, and $\kappa$ are the thermal expansion coefficient, the kinematic viscosity, and the thermal diffusivity of the fluid respectively. The above equations are nondimensionalized using $d$, $\Delta$, and $\kappa \sqrt{\mathrm{Ra}}/d$ as the length, temperature, and velocity scales respectively. For $\mathrm{Pr} = \infty$, the left hand side of Eq.~(\ref{eq:u}) vanishes, resulting in a linear equation~\cite{Pandey:PRE2014, Pandey:Pramana2016, Verma:POF2015} 
\begin{equation}
-\nabla \sigma + \theta \hat{z} + \frac{1}{\sqrt{\mathrm{Ra}}} \nabla^2 {\bf u} = 0, \label{eq:u_inf}
\end{equation}
which can be utilized to compute the Fourier modes of the velocity field from the corresponding modes of the temperature field~\citep{Pandey:PRE2014, Verma:POF2015}. Thus for $\mathrm{Pr} = \infty$, we solve Eqs.~(\ref{eq:u_inf}), (\ref{eq:T}), and (\ref{eq:m}) using a pseudospectral solver {\sc Tarang}~\cite{Verma:Pramana2013} in a two-dimensional square box. Stress-free boundary condition for the velocity field is employed on all the walls. Top and bottom plates are isothermal and sidewalls are adiabatic. Fields are dealiased using 2/3 rule. To satisfy the boundary conditions, velocity and temperature fields are expanded using free-slip basis functions~\cite{Verma:POF2015} as
\begin{eqnarray}
u_x(x,z) & = & \sum_{k_x,k_z} 4\hat{u}_x(k_x, k_z) \sin(k_x x) \cos(k_z z), \label{eq:basis_ux} \\
u_z(x,z) & = & \sum_{k_x,k_z} 4\hat{u}_z(k_x, k_z) \cos(k_x x) \sin(k_z z),	 \label{eq:basis_uZ} \\
\theta(x,z) & = & \sum_{k_x,k_z} 4\hat{\theta}(k_x, k_z) \cos(k_x x) \sin(k_z z), \label{eq:basis_th}
\end{eqnarray}
where $\hat{f}(k_x,k_z)$ represents the Fourier transform of a function $f(x,z)$. 

Researchers have tried to construct low dimensional models to mimic the dynamics of LSC reversals~\cite{Araujo:PRL2005, Brown:POF2008, Podvin:JFM2015, Ni:JFM2015}. Recently, \citet{Mannattil:EPJB2017} utilized the techniques of nonlinear time series analysis to conclude that the reversals in infinite-Pr RBC are however high dimensional. Therefore direct numerical simulations are important to prudently study the dynamics of reversals in the infinite-Pr RBC. Moreover, for very large Prandtl number RBC, the large- and small-scale quantities exhibit very similar scalings in two and three dimensions~\citep{Schmalzl:EPL2004, Pandey:Pramana2016}. Consequently, very large Prandtl number RBC can be studied in two dimensions, where very long time statistics are accessible at lower computational costs. Therefore, we integrated the governing equations for $\mathrm{Pr} = \infty$ and $\mathrm{Ra} = 10^8$ for a total time $t_{\mathrm{total}} = 3.52 \times 10^5 \, t_f$, where $t_f = \, d^2/(\kappa \sqrt{\mathrm{Ra}})$ is the unit time in the present case. A few important parameters of the simulation are summarized in Table~\ref{table:details}. We checked the resolution criterion by computing the time-averaged Batchelor length scale $\eta$, and find that its product with the largest wavenumber, $k_{\mathrm{max}} \eta \approx 2.6$, which indicates that the smallest length scales are properly resolved in our simulation. Note that the Batchelor length scale $\eta$ and the Kolmogorov length scale $\eta_K$ are related as $\eta = \eta_K/\sqrt{\mathrm{Pr}}$. Therefore, one needs to properly resolve the Batchelor scale in RBC with $\mathrm{Pr} > 1$~\cite{Chilla:EPJE2012, Shishkina:NJP2010}. Moreover, we compared the Nusselt number computed using the correlation of $u_z$ and $\theta$~\cite{Verma:NJP2017} with that using the exact relations derived from the Boussinesq equations~\cite{Shraiman:PRA1990}, and observed that they match within $1\%$, thus again indicating that our simulation is adequately resolved.
\begin{table}
\begin{ruledtabular}
\caption{Important parameters of the direct numerical simulation. $N^2$ is the total number of equidistant grid points in the simulation domain, $u_{\mathrm{rms}}$ is the root mean square velocity, computed as $\sqrt{ \langle u_x^2 + u_z^2 \rangle_{A,t}}$, where $\langle \cdot \rangle_{A,t}$ represents averaging over the entire simulation domain and time.}
\begin{tabular}{cccccc}
$\mathrm{Pr}$ & $\mathrm{Ra}$ & $N^2$ & Simulation time & $u_{\mathrm{rms}}$ & $k_{\mathrm{max}} \eta$ \\
\hline
$\infty$ & $10^8$ & $256^2$ & $3.52 \times 10^5 \, t_f$ & $1.42$ & 2.6
\end{tabular}
\label{table:details}
\end{ruledtabular}
\end{table}

%%%%%%%%%%%%%%%%%%%%%%%%%%%%%%%%%%%%%%%%%%%%%%%%%%%%%%%%%%%%%%%%%%%%%%%%%%%%%%%%%%%%%%%%
%%%%%%%%%%%%%%%%%%%%%%%%%%%%%%%%%%%%%%%%%%%%%%%%%%%%%%%%%%%%%%%%%%%%%%%%%%%%%%%%%%%%%%%%

\section{Results}
To explore the dynamics and statistical properties of reversals, we recorded the time history of the velocity and temperature fields at various locations in our simulation domain. Additionally, to understand the mechanism of reversals we tracked the temporal evolution of some of the most energetic Fourier modes of the flow.

\subsection{Statistical properties of reversals} \label{subsec:rev_stat}
In this subsection, we discuss the statistics of reversals using the velocity field monitored near the left sidewall. 

\begin{figure}
\includegraphics[scale=0.2]{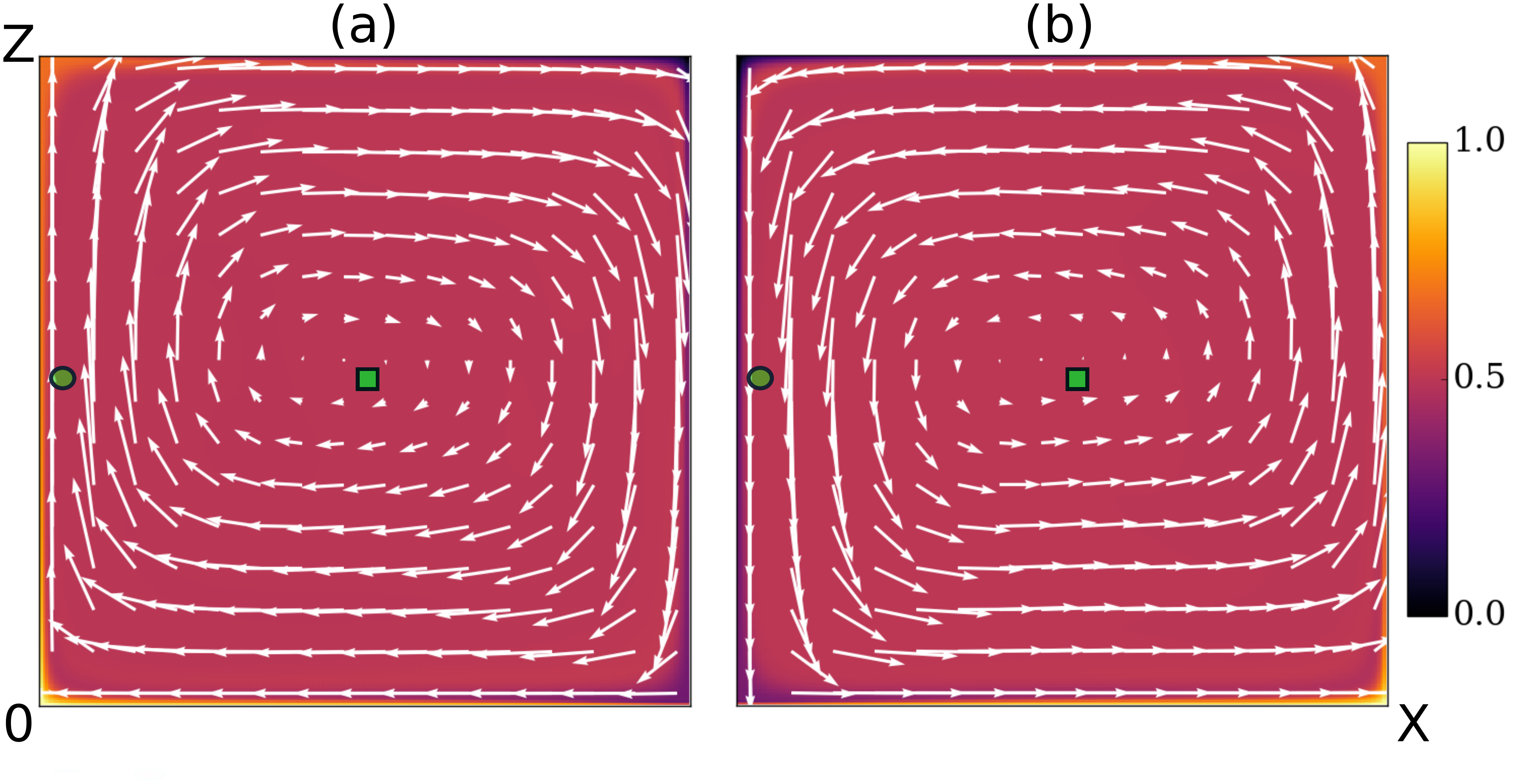}
\caption{Two stable flow configurations of the convective flow in a 2D square box with stress-free walls: LSC in the clockwise direction (a) and in the counterclockwise direction (b). Temperature field is shown as density plots and velocity field is represented by vectors. Time evolution of the temperature and velocity fields are tracked for the entire duration of simulation at the two probes located near the left wall (green circles) and at the center (green squares).}
\label{fig:vel_temp}
\end{figure}

After reaching the statistically steady state, we continued our simulation for a very long time, and observe that the stable convective structure is a single circulating roll occupying the whole box. Figure~\ref{fig:vel_temp}(a) shows a stable structure  circulating in the clockwise direction. As the flow evolves, this large-scale circulation (LSC) persists its clockwise direction for some time before switching its motion in the counterclockwise direction -- the other stable configuration, as exhibited in Fig.~\ref{fig:vel_temp}(b). The flow structure keeps oscillating between these two stable configurations during our entire observation time. However during the reversal events, flow structure becomes more complex. Multi-cell patterns dominate during reversals~\cite{Breuer:EPL2009, Petschel:PRE2011, Chandra:PRE2011, Chandra:PRL2013, Verma:POF2015}. In other words, higher-wavenumber Fourier modes become active and dominate during the flow reversals. We show a sequence of flow patterns during a reversal event in the Supplementary Video~\citep{reversal_movie}.
We track the vertical velocity $u_z(t)$ at two different probes in the simulation domain. One probe is located near the center of the left wall (indicated in Fig.~\ref{fig:vel_temp} by green circles) at $(x = 0.0625, z = 0.5)$, and henceforth will be referred to as the left probe (LP). The other probe, located at the center of the box (indicated by green squares in Fig.~\ref{fig:vel_temp}) will be referred to as the center probe (CP). 

\begin{figure}
\includegraphics[scale=0.35]{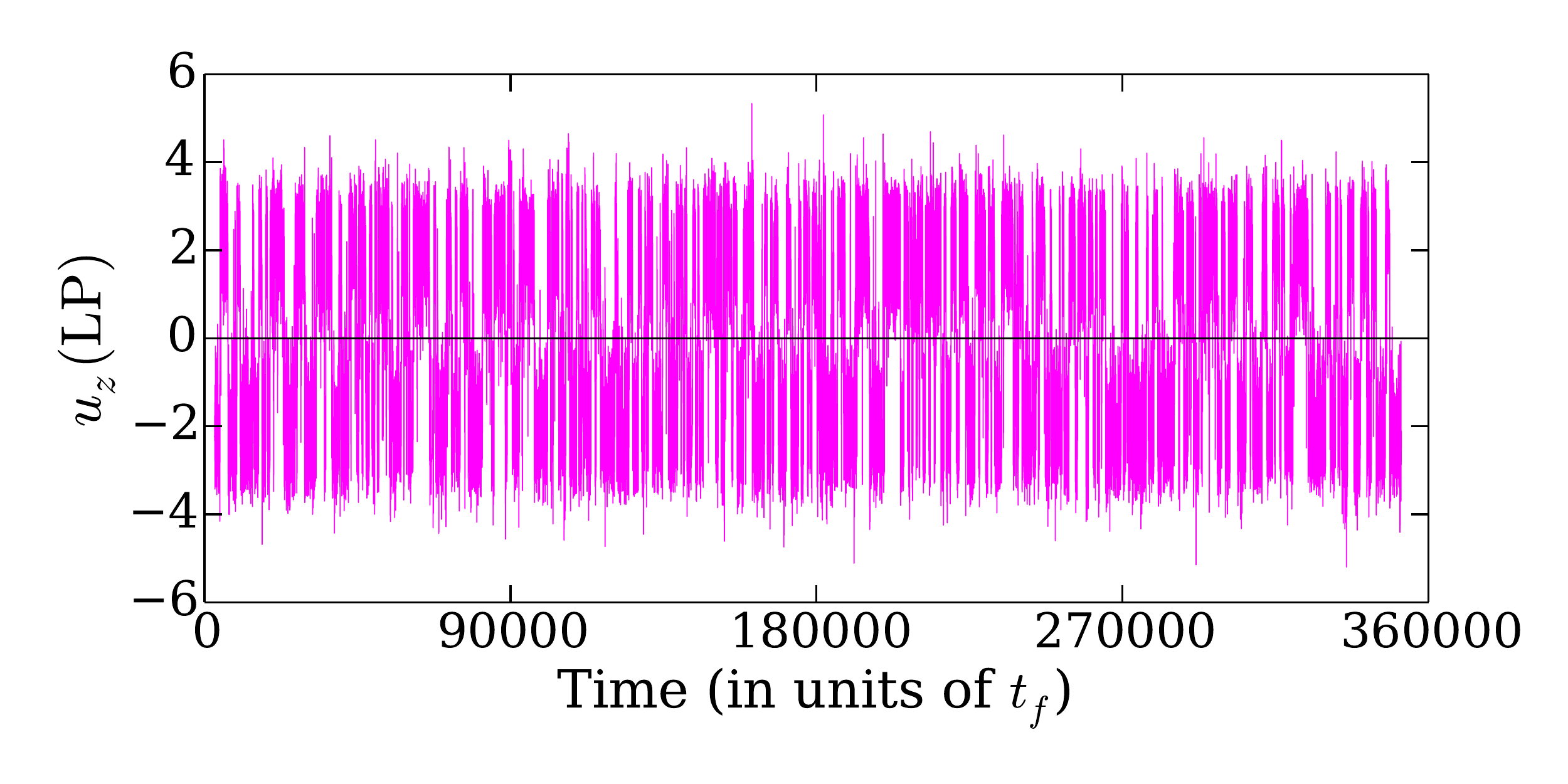}
\caption{Evolution of the vertical velocity at the left probe (indicated in Fig.~\ref{fig:vel_temp} by green circles). Velocity changes its sign irregularly, indicating occurrence of flow reversals.}
\label{fig:time_uz}
\end{figure}

\begin{figure}
\includegraphics[scale=0.35]{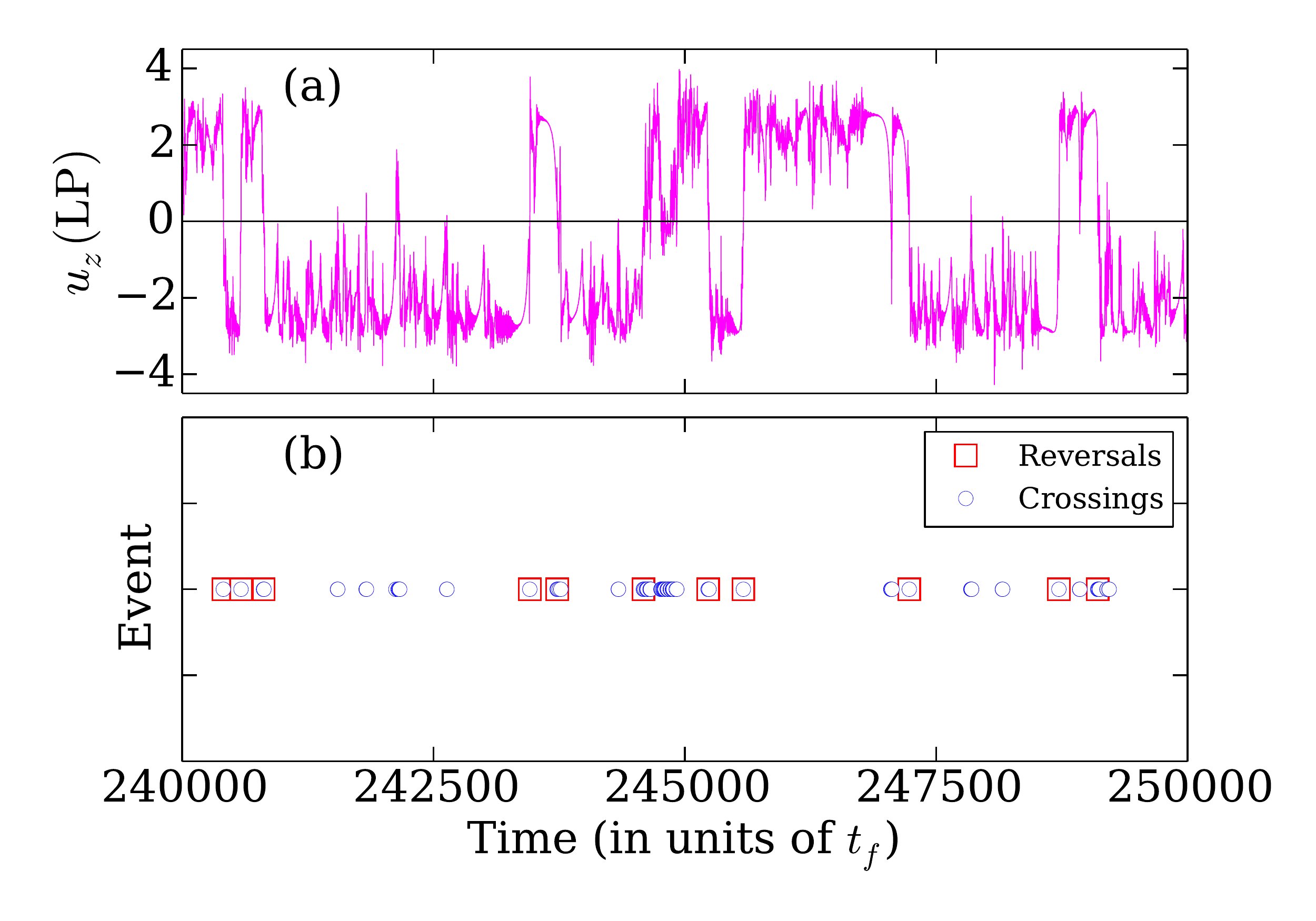}
\caption{(a) Magnification of Fig.~\ref{fig:time_uz} showing a few reversal and crossing events, which are indicated in panel (b).}
\label{fig:rev_cross}
\end{figure}

Figure~\ref{fig:time_uz} exhibits the time series of $u_z(\mathrm{LP})$ for the whole duration of simulation. It is evident from the figure that the vertical velocity at the left probe switches sign irregularly, indicating that the flow reverses repeatedly during our simulation. The velocity component $u_z(\mathrm{LP})$ fluctuates around a non-zero mean value between any two reversals. Each occurrence of the sign change of $u_z(\mathrm{LP})$ is termed as a ``crossing" event.
It is important to note however that all the crossings do not lead to flow reversals, as some of them might occur due to momentary decay of the primary flow mode, and (at the same time) the growth of the secondary modes~\citep{Mishra:JFM2011, Chandra:PRL2013}. All reversals however are crossings. This is illustrated in Fig.~\ref{fig:rev_cross}(a) where we show the temporal evolution of $u_z(\mathrm{LP})$ on an extended scale, which exhibits a few reversal and crossing events. Figure~\ref{fig:rev_cross}(b) indicates the instances of the occurrence of reversal and crossing events in Fig.~\ref{fig:rev_cross}(a). We distinguish reversals from crossings by putting a constraint on the waiting time between two consecutive crossings; a crossing is counted as a reversal only if it is separated from its neighboring crossings by at least $40 \, t_f$.

\begin{figure}
\includegraphics[scale=0.4]{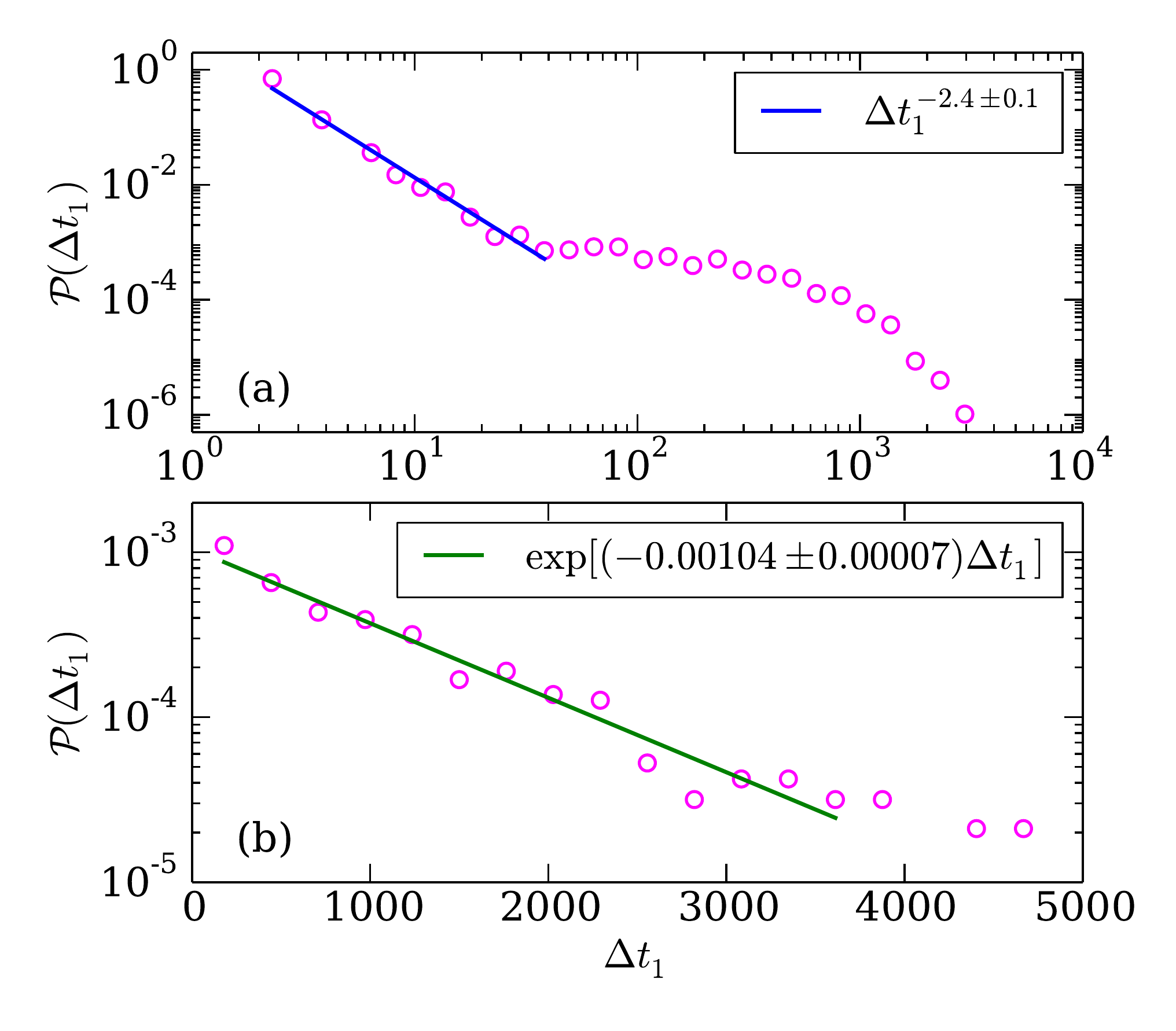}
\caption{(a) Probability distribution function of the waiting time between two consecutive crossings $\mathcal{P}(\Delta t_1)$ on a double logarithmic scale, which exhibits a powerlaw distribution for the shorter waiting times. (b) Probability distribution of the waiting time between two consecutive reversals however follows an exponential distribution.}
\label{fig:pdf_wait}
\end{figure}

By counting the number of sign changes in the time series of $u_z(\mathrm{LP})$, we observe 2591 crossings during the entire observation time. Moreover, crossings occur on a fast time scale compared to the mean waiting time between two consecutive flow reversals. Let us denote $t_n$ as the time when $n^{th}$ crossing occurs~\citep{Sreenivasan:PRE2002}, and the time gap between $n^{th}$ and $(n+r)^{th}$ crossings as $\Delta t_r$. The probability distribution function (PDF) of the waiting time between any two consecutive crossings, $\mathcal{P}(\Delta t_1)$, is exhibited in Fig.~\ref{fig:pdf_wait}(a) on a double logarithmic scale. It is evident that the PDF decays as a powerlaw for shorter waiting times, with best fit yielding $\mathcal{P}(\Delta t_1) \sim \Delta t_1^{-2.4}$. The powerlaw decay of $\mathcal{P}(\Delta t_1)$ for shorter waiting times has also been observed in experiments with moderate Prandtl number fluids by \citet{Sreenivasan:PRE2002, Xi:PRE2006}, and \citet{Brown:JFM2006}, albeit with lower scaling exponents around one. The larger exponent observed here might be due to two dimensionality or due to very large Prandtl number of the flow.

One can observe in Fig.~\ref{fig:pdf_wait}(a) that the powerlaw region extends only up to $\Delta t_1 \approx 40 \, t_f$, whereas some other scaling holds for larger waiting times. We find that $\mathcal{P}(\Delta t_1)$ decays  exponentially for $\Delta t_1 \gtrsim 40 \, t_f$. It has been observed in experiments~\cite{Sreenivasan:PRE2002, Xi:PRE2006, Brown:JFM2006, Xi:PRE2007, Huang:JFM2016} and two-dimensional numerical simulations~\cite{Podvin:JFM2015} of moderate Prandtl number RBC that the waiting times between two consecutive LSC reversals follow a Poissonian distribution. \citet{Podvin:JFM2015} detected flow reversals by tracking sign changes in the global angular momentum of their 2D flow, which correspond to occurrence of flow reversals. Thus it can be inferred that the exponential distribution of $\mathcal{P}(\Delta t_1)$ for $\Delta t_1 \gtrsim 40 \, t_f$ occurs due to reversal events, and that these reversals are separated from crossings by a separation time scale $t_s \approx 40 \, t_f$~\citep{Brown:JFM2006}. Therefore as mentioned earlier, to count the number of true reversal events, we discarded crossings which are separated from their neighboring crossing events by less than $40 \, t_f$. Using this criterion, we find only 359 reversal events compared to 2591 crossings in our time series of $u_z(\mathrm{LP})$. The mean waiting time between two consecutive reversals is $(\Delta t_1)_\mathrm{mean} \approx 975 \, t_f$. 

In Fig.~\ref{fig:pdf_wait}(b) we plot the PDF of waiting times between two consecutive reversals on a semilogarithmic scale, which exhibits that $\mathcal{P}(\Delta t_1)$ for reversals can indeed be fitted well by an exponential distribution. The best fit yields $\mathcal{P}(\Delta t_1) \sim \exp[(-0.00104 \pm 0.000066) \Delta t_1]$, suggesting a mean waiting time between two consecutive reversals as $(\Delta t_1)_\mathrm{mean} \approx (962 \pm 60) \, t_f$, which is consistent with the aforementioned $(\Delta t_1)_\mathrm{mean} \approx 975 \, t_f$.  For $\Delta t_1 \gtrsim 3700 \, t_f$, $\mathcal{P}(\Delta t_1)$ shows deviations from the exponential behavior, which is due to insufficient statistics for very long waiting times. Note that in our simulation the circulation time of LSC is $t_c \approx 4/u_{\mathrm{rms}} \approx 4/1.4 \approx 2.8 \, t_f$, thus the reversals occur at every $350 \, t_c$.

\begin{figure}
\includegraphics[scale=0.33]{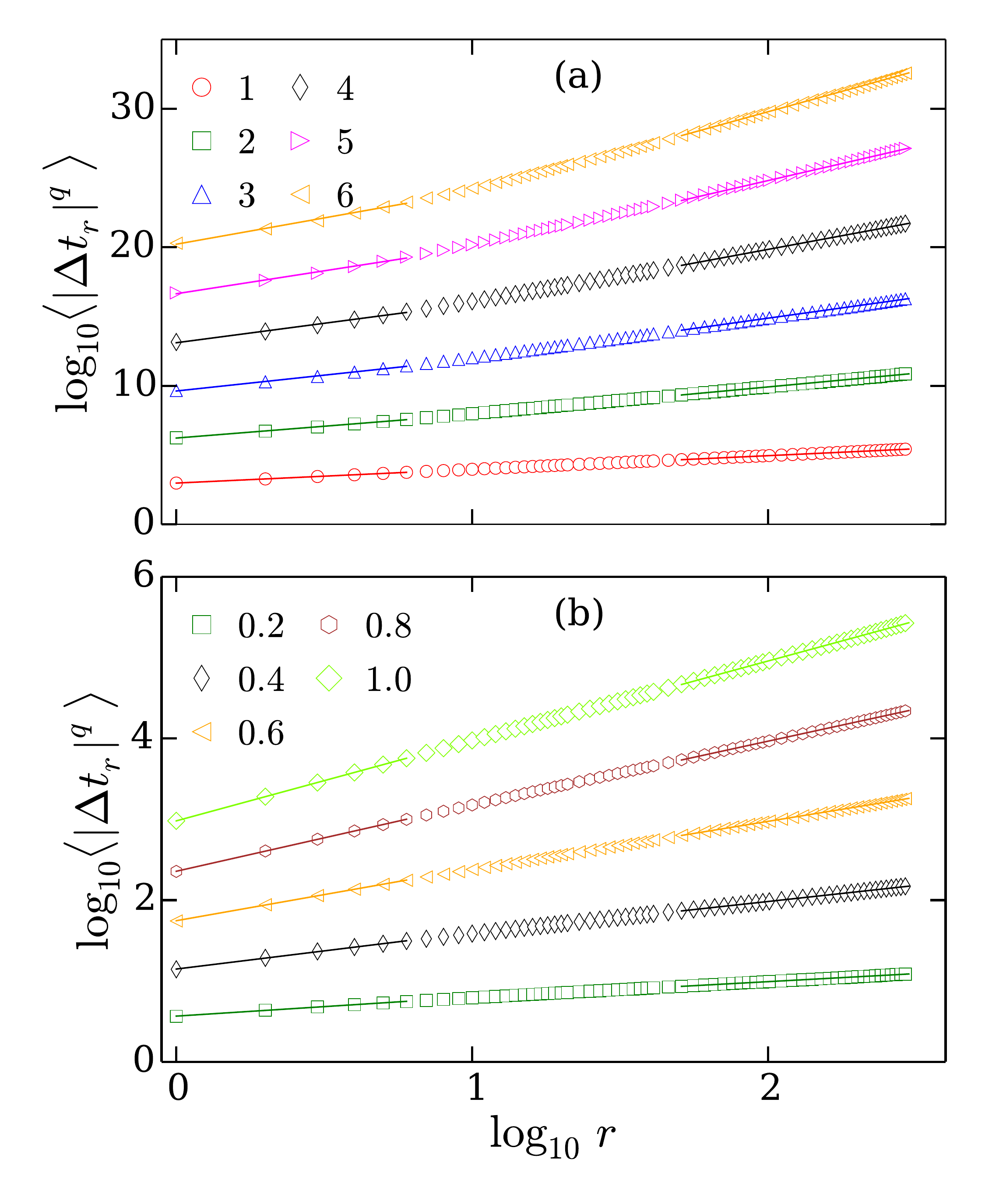}
\caption{Moments of the generalized interswitch interval $\langle |\Delta t_r|^q \rangle$ between reversals as function of $r$ for (a) $q = 1$ to 6 and (b) $q \leq 1$ ($q$ increases from bottom to top). We observe scaling regions for $r \leq 6$ and for $r \geq 50$.}
\label{fig:sf_rev}
\end{figure}

\begin{figure}
\includegraphics[scale=0.35]{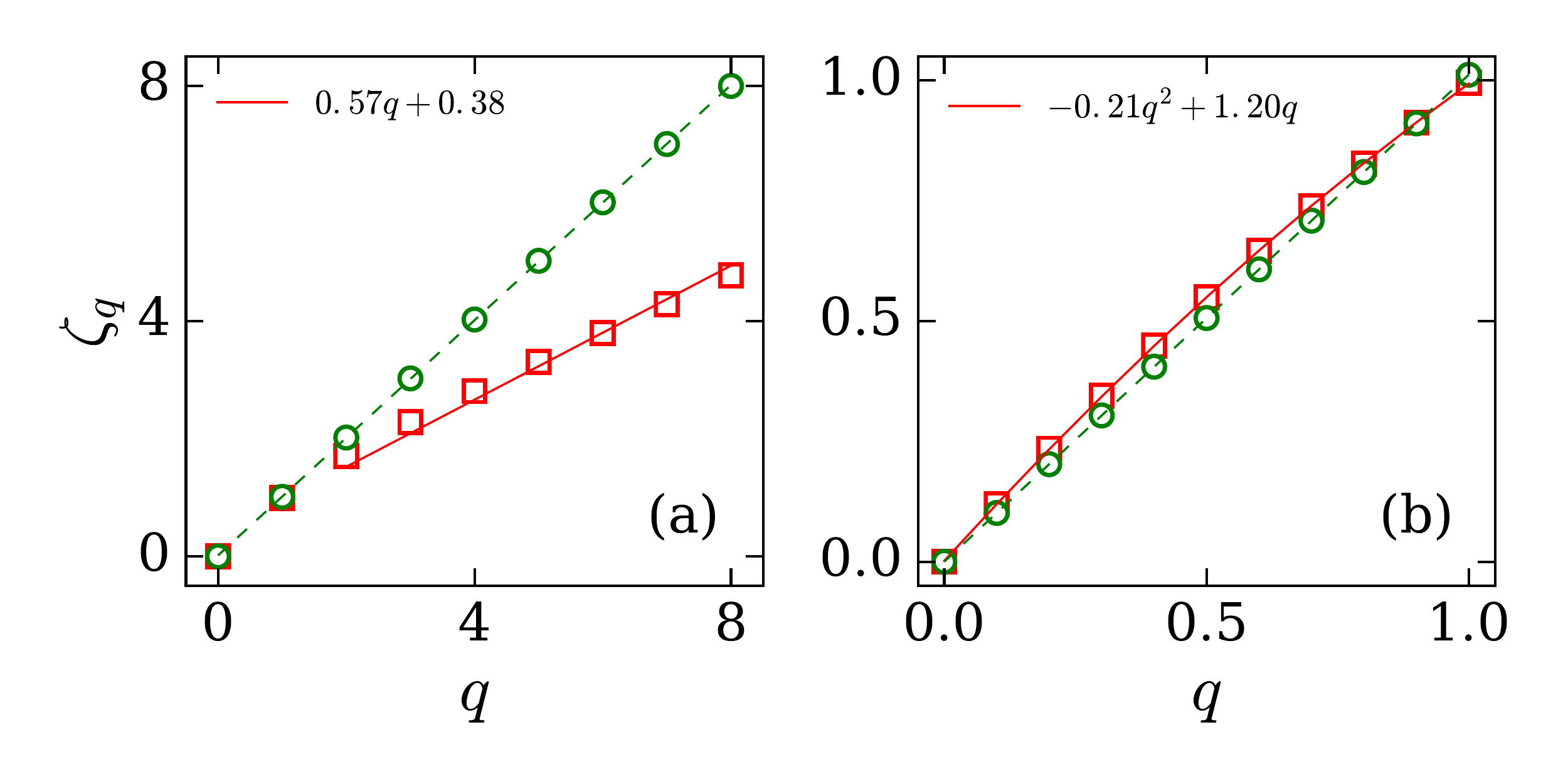}
\caption{Scaling exponents $\zeta_q$ as function of the order $q$ of the moments of generalized interswitch interval $\Delta t_r$ between reversals for (a) $q \geq 1$ and (b) $q \leq 1$. Exponents obtained for $r \leq 6$ (red squares) scale linearly for $q \geq 1$, whereas show a very weak nonlinear behavior for $q \leq 1$. However for $r \geq 50$, the exponents scale as $\zeta_q \cong q$ (green circles), revealing a decorrelation between distant reversals.}
\label{fig:exp_sf_rev}
\end{figure}

Following \citet{Sreenivasan:PRE2002}, we compute the moments of generalized interswitch intervals defined as $\langle |\Delta t_r|^q \rangle$ to gain further insight. Here $\Delta t_r = |t_{n+r}-t_n|$ is the time interval between $n^{th}$ and $(n+r)^{th}$ reversals. In Fig.~\ref{fig:sf_rev}(a), we plot the moments $\log_{10} \langle | \Delta t_r |^q \rangle$ for $q = 1$ to $6$ as function of $\log_{10} r$, where $\langle \cdot \rangle$ represents the running average along the entire time series. We find that the moments exhibit two scaling regimes, one for $r \leq 6$ and another for $r \geq 50$, with $\langle | \Delta t_r |^q \rangle \sim r^{\zeta_q}$ for these regimes. We compute $\zeta_q$ using the least square fit, and plot them as a function of $q$ in Fig.~\ref{fig:exp_sf_rev}(a).  We find $\zeta_q \cong q$ for $r \geq 50$, which suggests that the distant reversals are decorrelated. For $r \leq 6$ however $\zeta_q \cong 0.57 q + 0.38$, suggesting a correlation between neighboring reversals, which has been credited to occur due to finite-size effect~\citep{Sreenivasan:PRE2002}.

We also compute the moments of $\Delta t_r$  for $q < 1$, and plot them in Fig.~\ref{fig:sf_rev}(b). Here too, we find that $\zeta_q$ increases linearly with $q$ for $r \geq 50$, again showing a decorrelation of the distant reversal events [see Fig.~\ref{fig:exp_sf_rev}(b)]. For $r \leq 6$ however $\zeta_q$ exhibits a weakly nonlinear behavior as shown in Fig.~\ref{fig:exp_sf_rev}(b), following $\zeta_q = -0.21 q^2 + 1.20 q$. 

%%%%%%%%%%%%%%%%%%%%%%%%%%%%%%%%%%%%%%%%%%%%%%%%%%%%%%%%%%%%%%%%%%%%%%%%%%%%%%%%%%%%%%%%
%%%%%%%%%%%%%%%%%%%%%%%%%%%%%%%%%%%%%%%%%%%%%%%%%%%%%%%%%%%%%%%%%%%%%%%%%%%%%%%%%%%%%%%%

\subsection{Statistical properties of dominant flow modes} \label{subsec:modes}

\begin{figure}
\includegraphics[scale=0.29]{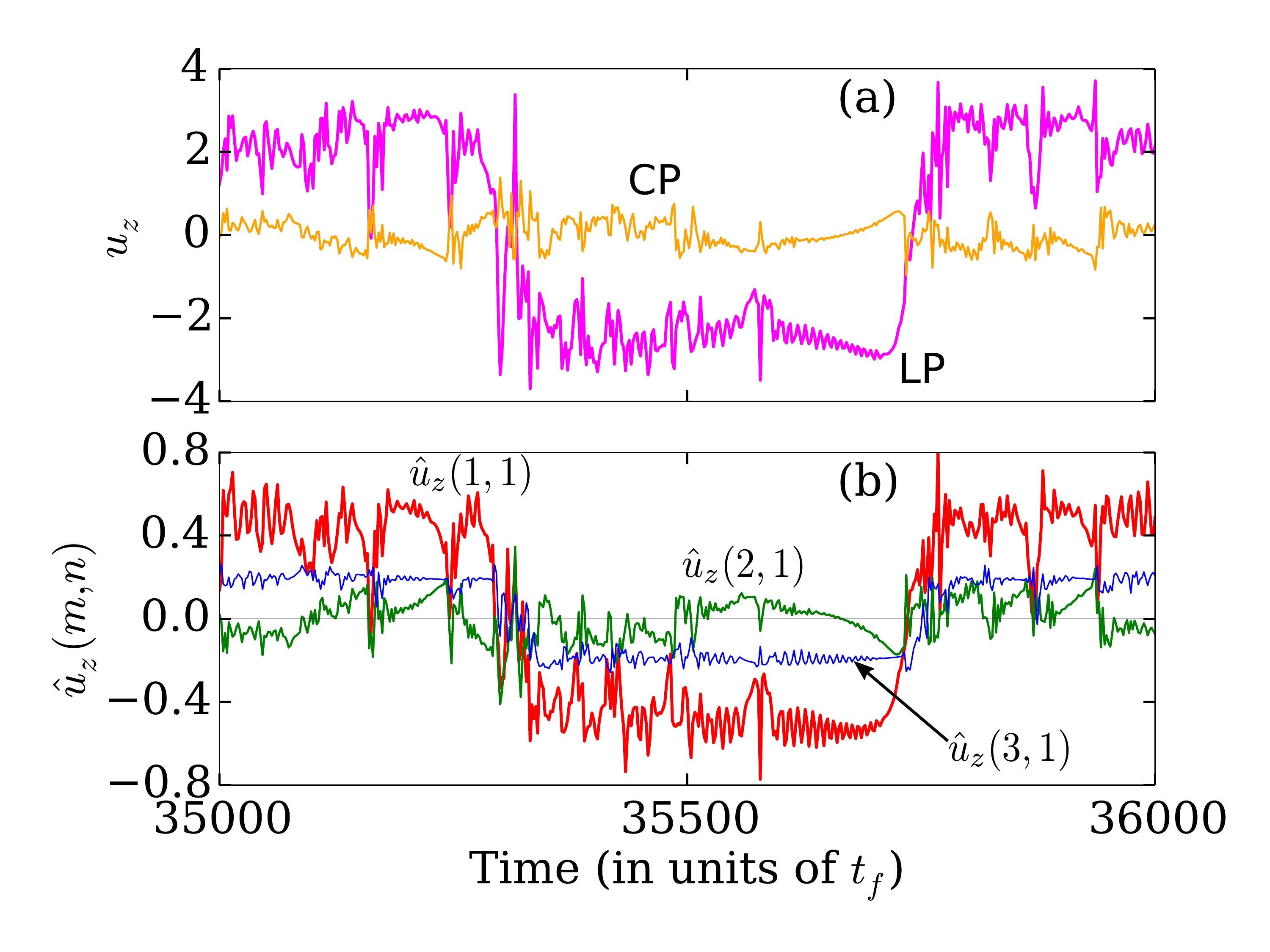}
\caption{(a) Vertical velocities at the left and the center probes as function of time, (b) and the evolution of the most dominant Fourier modes during the same time interval. A fraction of the full time series covering two reversals is shown here.}
\label{fig:time_modes}
\end{figure}

As mentioned earlier, we also record the temporal evolution of vertical velocity at the center of our simulation domain, $u_z(\mathrm{CP})$, as well as the amplitudes of the most energy containing Fourier modes. In Fig.~\ref{fig:time_modes}(a) we plot the evolution of $u_z(\mathrm{LP})$ and $u_z(\mathrm{CP})$ covering two consecutive reversals. We observe that $u_z(\mathrm{LP})$ fluctuates around two nonzero mean values most of the time, but $u_z(\mathrm{CP})$ fluctuates around zero. Moreover, we find that $u_z(\mathrm{CP})$ is anticorrelated with $u_z(\mathrm{LP})$ by computing their cross-correlation, defined as $\langle u_z(\mathrm{CP}, t+\tau) u_z(\mathrm{LP}, t) \rangle$, and getting negative values for time lags $\tau \lesssim 80 \, t_f$. This anticorrelation between velocities at the left and center probes is similar to those observed between the primary and secondary flow modes during a reversal or cessation event as observed in \citet{Mishra:JFM2011, Petschel:PRE2011, Chandra:PRE2011, Chandra:PRL2013}, and \citet{Verma:POF2015}. 

\begin{figure}
\includegraphics[scale=0.62]{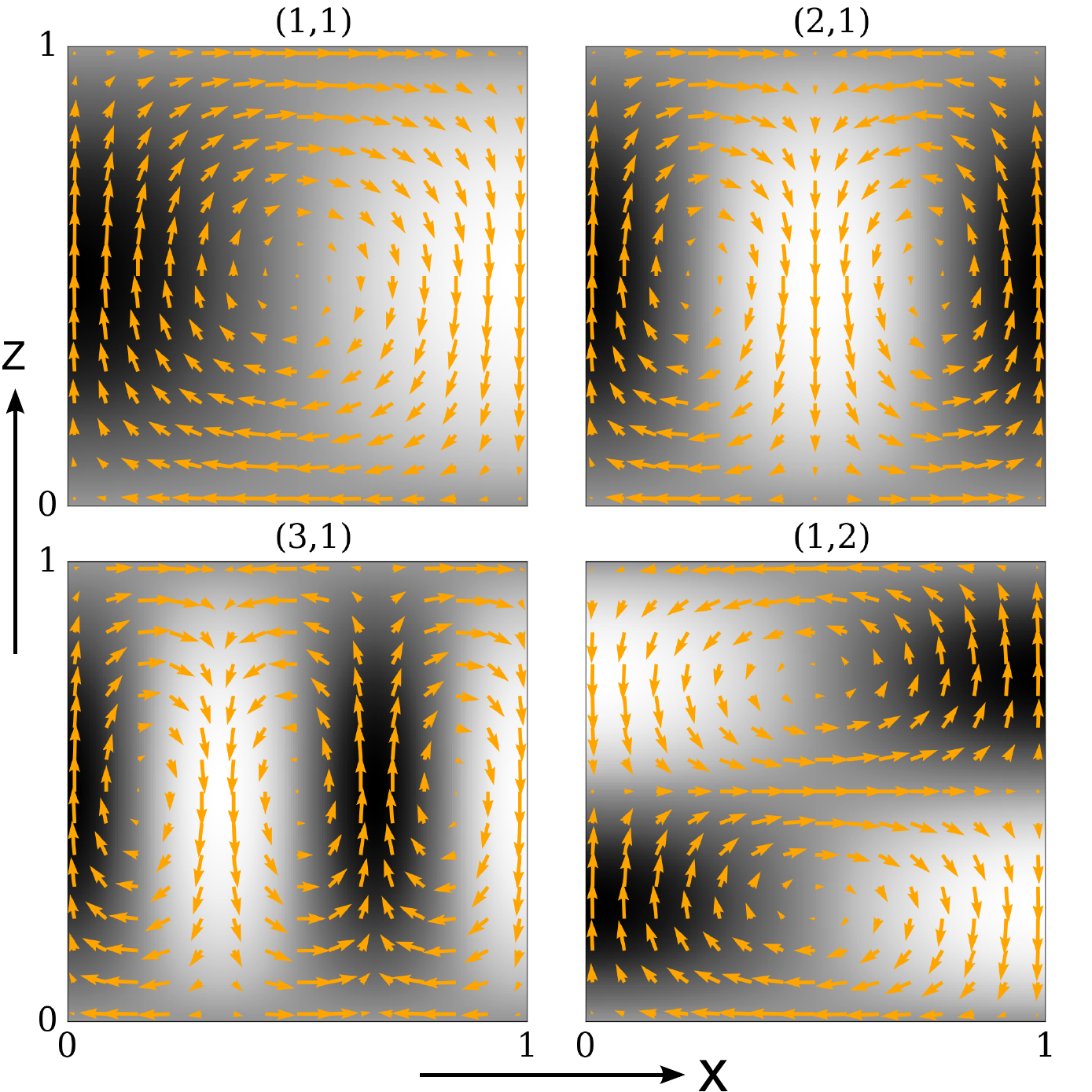}
\caption{Velocity (arrows) and temperature fluctuation fields (grey scale) corresponding to Fourier modes with indices $(m,n)$ computed using Eqs.~(\ref{eq:basis_ux}--\ref{eq:basis_th}). Dark and bright regions represent the hottest and coldest fluids respectively.}
\label{fig:Fourier_modes}
\end{figure}

In Fig.~\ref{fig:time_modes}(b) we plot the amplitudes of the Fourier modes $\hat{u}_z(1,1)$, $\hat{u}_z(2,1)$, and $\hat{u}_z(3,1)$ for the same time interval as in Fig.~\ref{fig:time_modes}(a). These are the most energetic modes (see Table~IV in \citet{Verma:POF2015}). Note that the wavenumber components for a Fourier mode with indices $(m,n)$ are $k_x = m\pi, \, k_z = n\pi$,  which represents a pattern with $m$-rolls along $x$-direction and $n$-rolls along $z$-direction. We illustrate a few low-wavenumber Fourier modes in Fig.~\ref{fig:Fourier_modes}, where we can see that the mode $(1,1)$ represents a single roll occupying the whole box. Similarly the modes $(2,1)$ and $(1,2)$ respectively represent two rolls stacked along $x$- and $z$-directions. It is evident from Fig.~\ref{fig:time_modes}(b) that the modes $\hat{u}_z(1,1)$, $\hat{u}_z(3,1)$, and $\hat{u}_z(2,1)$ capture the evolution of $u_z(\mathrm{LP})$ and $u_z(\mathrm{CP})$ very well. For instance, the modes $\hat{u}_z(1,1)$ and $\hat{u}_z(3,1)$ change their sign after the reversal events, as $u_z(\mathrm{LP})$ also does. 
\citet{Verma:POF2015} classified the Fourier modes according to their indices. In two dimensions, the modes belong to one of the four classes: odd-odd (OO), even-even (EE), even-odd (EO), and odd-even (OE). For instance, the modes $(1,1), (2,1), (1,2)$, and $(2,2)$ belong to OO, EO, OE, and EE classes respectively. Moreover, these four classes form an abelian group called {\it Klein four-group} $Z_2 \times Z_2$~\cite{Verma:POF2015}. Verma \textit{et al.}~\cite{Verma:POF2015} also deduced that in 2D RBC with free-slip walls, the OO modes switch their sign after a flow reversal, while the modes belonging to other classes do not. Moreover, the OE, EE, and EO modes fluctuate with their mean value around zero. We have analyzed four most dominant modes from each of the aforementioned classes, but for brevity, we focus only on $\hat{u}_z(1,1)$, $\hat{u}_z(2,1)$, and $\hat{u}_z(3,1)$.

\begin{figure}
\includegraphics[scale=0.35]{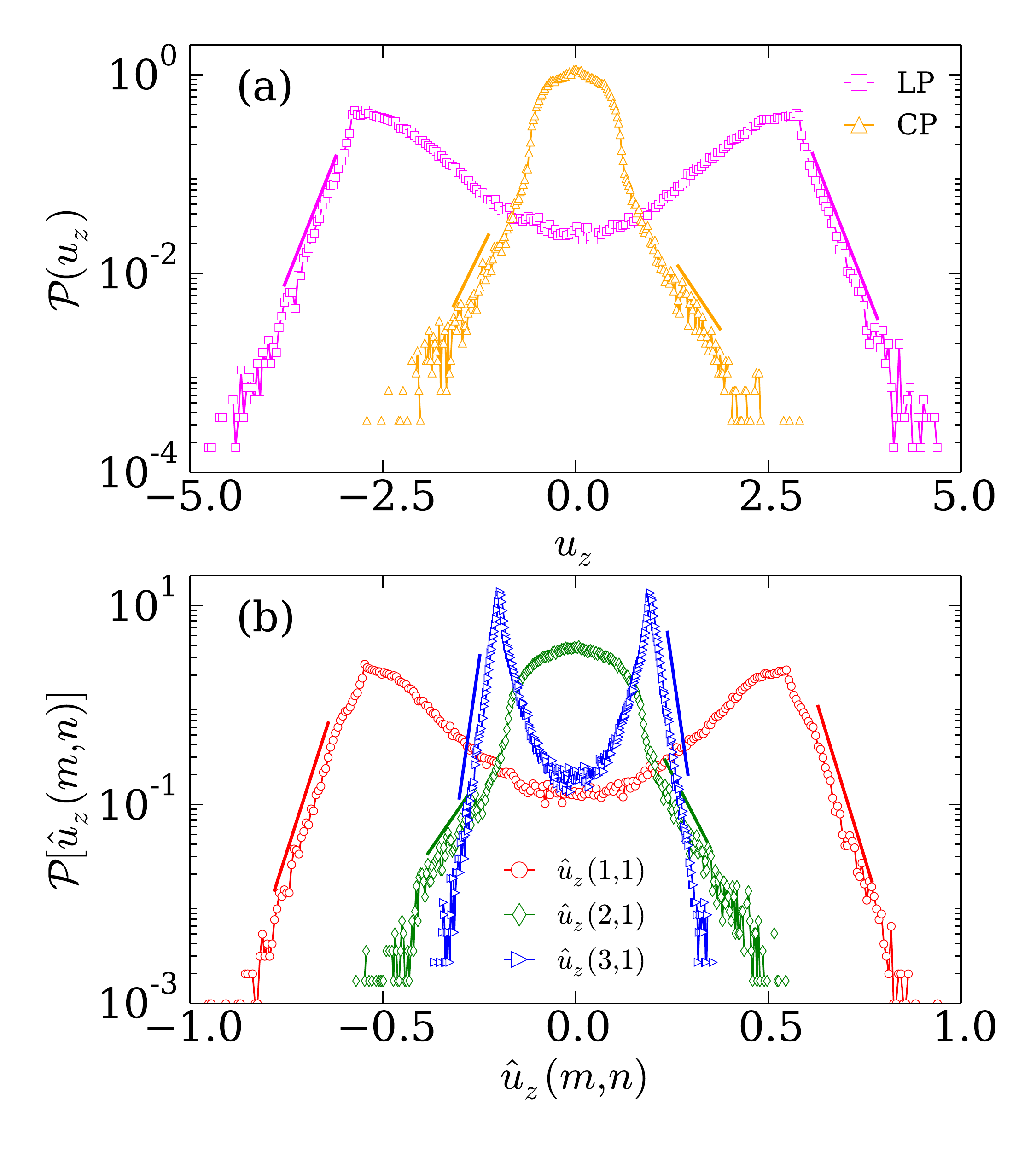}
\caption{Probability distribution function of the vertical velocities at the left and the center probes (a), and of the amplitudes of the dominant Fourier modes (b). The distribution is bimodal for $u_z(\mathrm{LP})$ and the OO modes, consistent with the fact that they change sign after flow reversals. PDFs of $u_z(\mathrm{CP})$ and $\hat{u}_z(2,1)$ exhibit broad peaks centered at zero and are non-Gaussian. The exponential tails are observed for all the distributions, and the exponential regions are indicated by straight lines.}
\label{fig:pdf_modes}
\end{figure}

As mentioned above, the OO modes change their sign after a flow reversal. For instance, we detect 2815, 2657 and 2715 crossings for $\hat{u}_z(1,1)$, $\hat{u}_z(3,1)$, and $\hat{u}_z(5,1)$ modes respectively, which is  close to 2367 crossings for $u_z(\mathrm{LP})$ during the same time interval. Similarly, we find that the probability distribution of waiting times $\Delta t_1$ for the OO modes (not shown here) are very similar to that for $u_z(\mathrm{LP})$ shown in Fig.~\ref{fig:pdf_wait}. In Fig.~\ref{fig:pdf_modes}(a), we plot the PDFs of $u_z(\mathrm{LP})$ and $u_z(\mathrm{CP})$, and find that $\mathcal{P}[u_z(\mathrm{LP})]$ is bimodal, which agrees with the fact that it fluctuates between two non-zero mean values. However, $\mathcal{P}[u_z(\mathrm{CP})]$ shows a Gaussian-like distribution, with a broad peak at zero. We find that the kurtosis of $\mathcal{P}[u_z(\mathrm{CP})]$ is 4.7, which indicates that it deviates from the Gaussian distribution, but not very strongly. We show the PDFs of $\hat{u}_z(1,1)$, $\hat{u}_z(2,1)$, and $\hat{u}_z(3,1)$ in Fig.~\ref{fig:pdf_modes}(b), and find that the distribution of the OO modes are bimodal, similar to $\mathcal{P}[u_z(\mathrm{LP})]$. Moreover, similar to $\mathcal{P}[u_z(\mathrm{CP})]$, $\mathcal{P}[\hat{u}_z(2,1)]$ also exhibits a Gaussian-like distribution with its kurtosis value nearly equal to 3.4, which reveals that the deviation from the Gaussian behavior is weak. This observation that the statistical properties of $\hat{u}_z(2,1)$ are similar to those of $u_z(\mathrm{CP})$ is not surprising since the vertical velocity at the center gets most dominant contribution from the $\hat{u}_z(2,1)$ Fourier mode (see Fig.~\ref{fig:Fourier_modes}). We also examined the PDFs of EE, OE, and other EO modes (not shown here), that exhibit strongly non-Gaussian behavior with their kurtosis values close to 10. Moreover, we find that the tails of all the PDFs are exponential.

We also observe from Fig.~\ref{fig:time_modes} that $u_z(\mathrm{LP})$ and the OO modes are autocorrelated for longer time compared to $u_z(\mathrm{CP})$ and the (2,1) mode. Therefore to quantify this, we compute the autocorrelation function for these time series as
\begin{equation}
C(\tau) = \frac{ \langle u(t+\tau) u(t) \rangle } { \langle u^2(t) \rangle},
\end{equation}
where $u(t)$ represents the evolution of any of the aforementioned time series. A useful quantity that can be computed from the correlation function is the integral time scale defined as
\begin{equation}
\mathscr{T} = \int_0^{t_\mathrm{total}} C(\tau) d\tau,
\end{equation}
which is a measure of how long a quantity is correlated with itself. We find $\mathscr{T} \approx 340 \, t_f$ for $u_z(\mathrm{LP})$ and the OO modes, whereas $\mathscr{T} \approx 20 \, t_f$ for $u_z(\mathrm{CP})$ and $\hat{u}_z(2,1)$. Thus the OO modes are autocorrelated for much longer time compared to the modes from the other classes.

\begin{figure}
\includegraphics[scale=0.35]{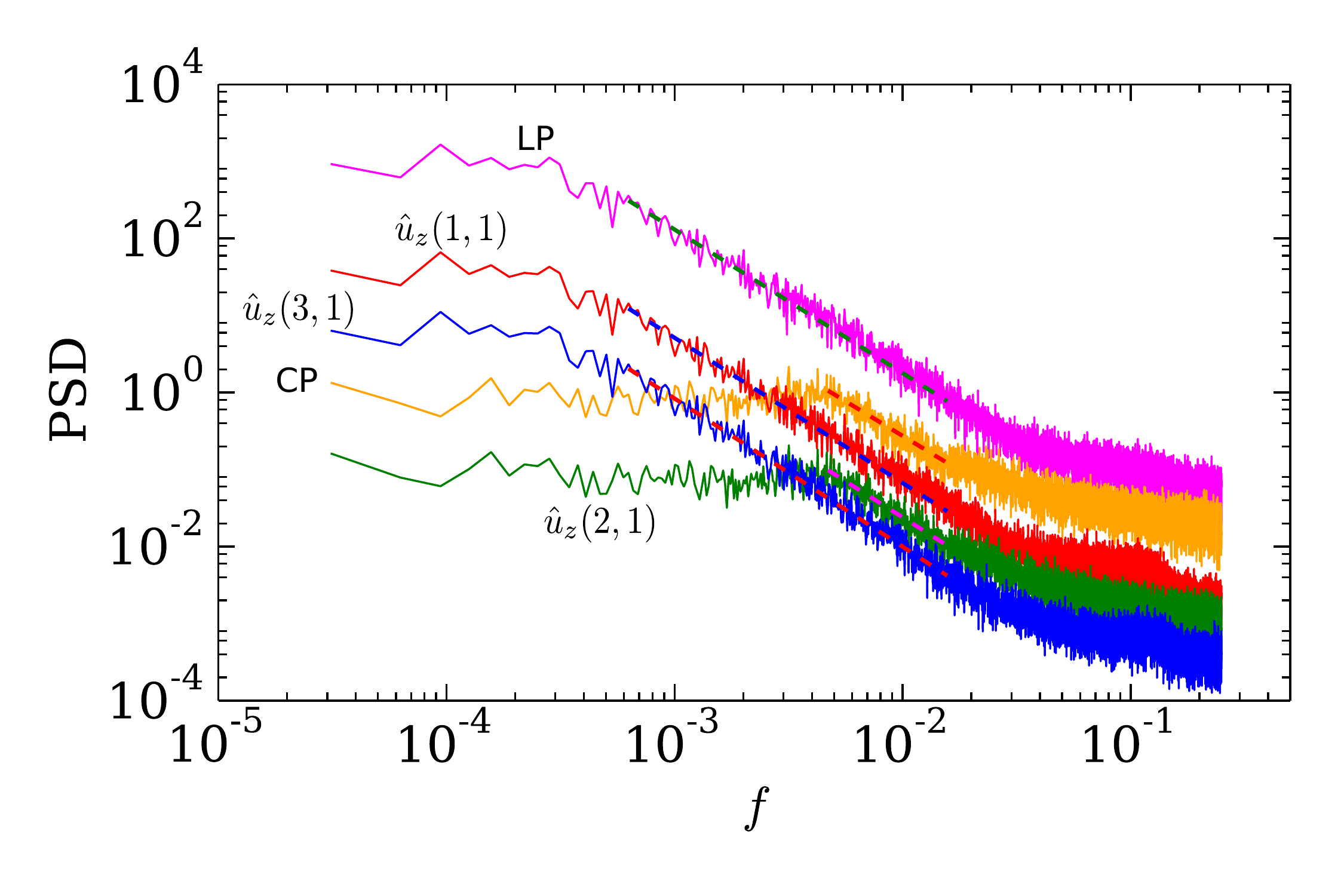}
\caption{Power spectral densities (PSD) of the vertical velocities at the left and the center probes, and of the Fourier modes. For the OO modes and $u_z(\mathrm{LP})$, we observe $\mathrm{PSD} \sim f^{-2}$ for a wide range of frequencies, whereas this scaling is observed only for a relatively narrower range of frequencies for $u_z(\mathrm{CP})$ and $\hat{u}_z(2,1)$.} 
\label{fig:fft_modes}
\end{figure}
To understand the physical process responsible for the evolution of these quantities, we compute their power spectral densities (PSD), and exhibit them in Fig.~\ref{fig:fft_modes}. It is apparent that the PSDs of $u_z(\mathrm{LP})$ and the OO modes show $1/f^{\alpha}$ scaling for a wide range of frequencies. The velocity components $u_z(\mathrm{CP})$ and $\hat{u}_z(2,1)$ show an equal power at small frequencies, but their PSDs decay as $1/f^{\alpha}$ too for a relatively narrower range of frequencies. Best fit to these regions yield $\alpha \approx 1.9 \pm 0.1$ for all the quantities, with the scaling ranges indicated in the figure by dashed lines. The $1/f^2$ power spectra of these signals are reminiscent of the Brownian process. The scaling range in time domain translates approximately from $10$ to $200 \, t_f$ for $u_z(\mathrm{LP})$ and the OO modes, and from 10 to $25 \, t_f$ for $u_z(\mathrm{CP})$ and $\hat{u}_z(2,1)$. The cutoff for the powerlaw behavior is close to the integral time scales for these quantities, whereas the lower cutoff scale is approximately of the order of a circulation time $t_c$ of LSC, with $t_c \approx 3 \, t_f$ in the present case. The $1/f^{\alpha}$ power spectra corresponding to long time fluctuations has been reported in other turbulent flows as well~\cite{Dmitruk:PRE2011}.
\begin{figure}
\includegraphics[scale=0.35]{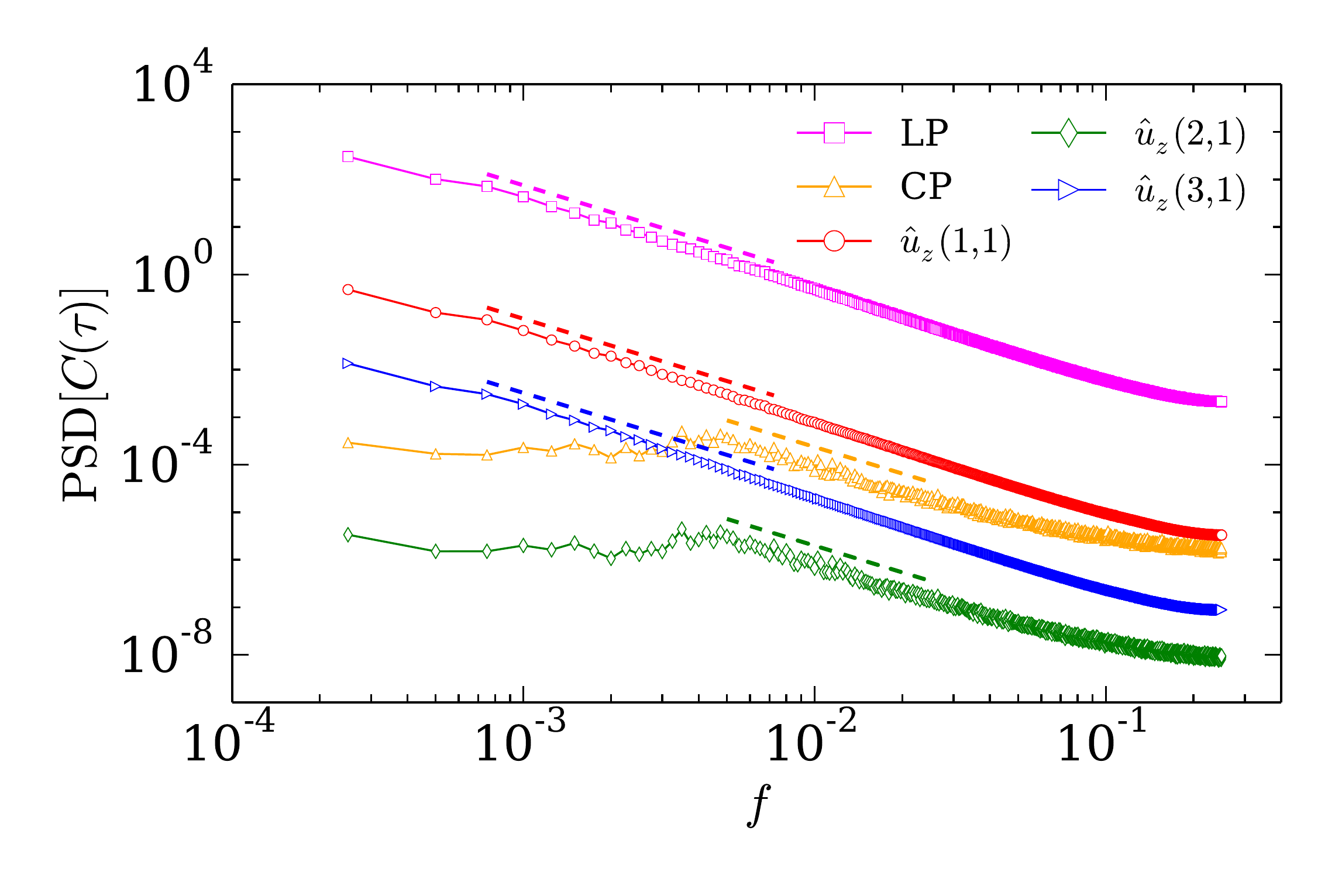}
\caption{Power spectral densities of the autocorrelation functions, $\mathrm{PSD}[C(\tau)]$, that also exhibit $f^{-2}$ scaling for similar range of frequencies as those in Fig.~\ref{fig:fft_modes}.}
\label{fig:fft_acor_modes}
\end{figure}

The power spectrum of the autocorrelation function is equivalent to the power spectrum of the signal. Therefore in Fig.~\ref{fig:fft_acor_modes} we plot the power spectra of the autocorrelation function, $\mathrm{PSD}[C(\tau)]$, of the aforementioned quantities. Using the least square fitting, we obtain a very similar $f^{-2}$ scaling for all the PSDs, with the scaling range remaining also very similar to those in power spectra of the original time series.

In the next subsection, we study the intermittency in the evolution of the aforementioned quantities by computing the temporal structure functions.

%%%%%%%%%%%%%%%%%%%%%%%%%%%%%%%%%%%%%%%%%%%%%%%%%%%%%%%%%%%%%%%%%%%%%%%%%%%%%%%%%%%%%%%%
%%%%%%%%%%%%%%%%%%%%%%%%%%%%%%%%%%%%%%%%%%%%%%%%%%%%%%%%%%%%%%%%%%%%%%%%%%%%%%%%%%%%%%%%

\subsection{Structure functions and intermittency} \label{subsec:str_fns}
\begin{figure}
\includegraphics[scale=0.35]{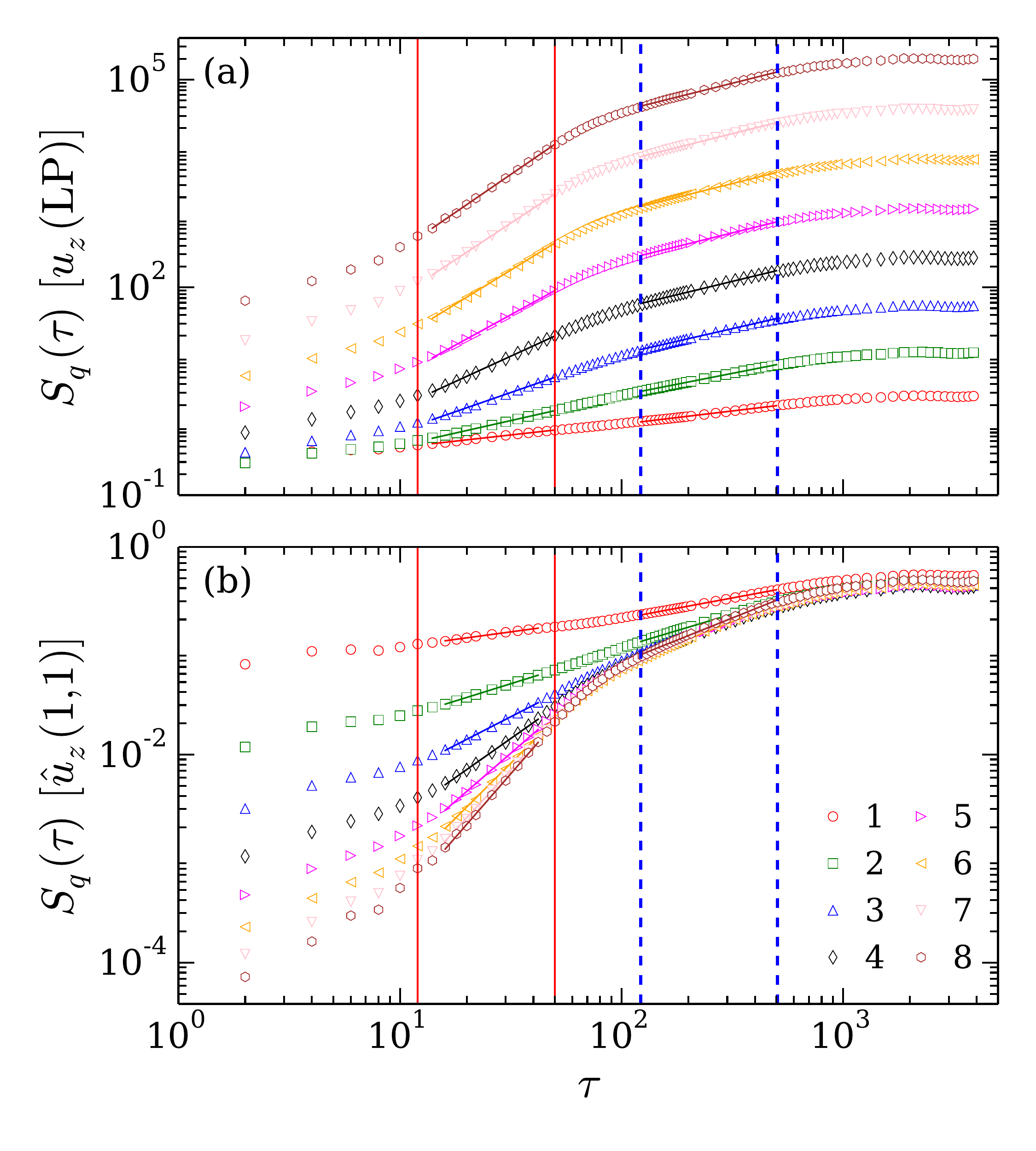}
\caption{Structure functions $S_q(\tau)$ for $q = 1$ to 8 for (a) $u_z(\mathrm{LP})$ and (b) $\hat{u}_z(1,1)$. Two scaling regimes located between red solid and blue dashed lines can be identified.}
\label{fig:SF}
\end{figure}

We compute the $q^{th}$ order structure function for a time series $u(t)$ as:
\begin{equation}
S_q(\tau) = \langle |u(t+\tau)-u(t)|^q \rangle.
\end{equation}
In Fig.~\ref{fig:SF}(a) and (b), we plot $S_q(\tau)$ for $u_z(\mathrm{LP})$ and $\hat{u}_z(1,1)$ respectively for $q$ = 1 to 8, and observe two different scaling regimes, where $S_q(\tau) \sim \tau^{\mu_q}$. The first scaling regime is observed for $10 \, t_f \lesssim \tau \lesssim 50 \, t_f$ (indicated between two vertical solid red lines), whereas the other scaling regime can be recognized for $100 \, t_f \lesssim \tau \lesssim 300 \, t_f$ (between two vertical dashed blue lines). Moreover, Fig.~\ref{fig:SF} exhibits that the structure functions saturate for $\tau \gtrsim 1000 \, t_f$, which is because $u_z(\mathrm{LP})$ and $\hat{u}_z(1,1)$ become decorrelated for $\tau \gtrapprox \mathcal{O}(\mathscr{T})$, with $\mathscr{T} \approx 350 \, t_f$ for $u_z(\mathrm{LP})$ and the OO modes. 

\begin{figure}
\includegraphics[scale=0.4]{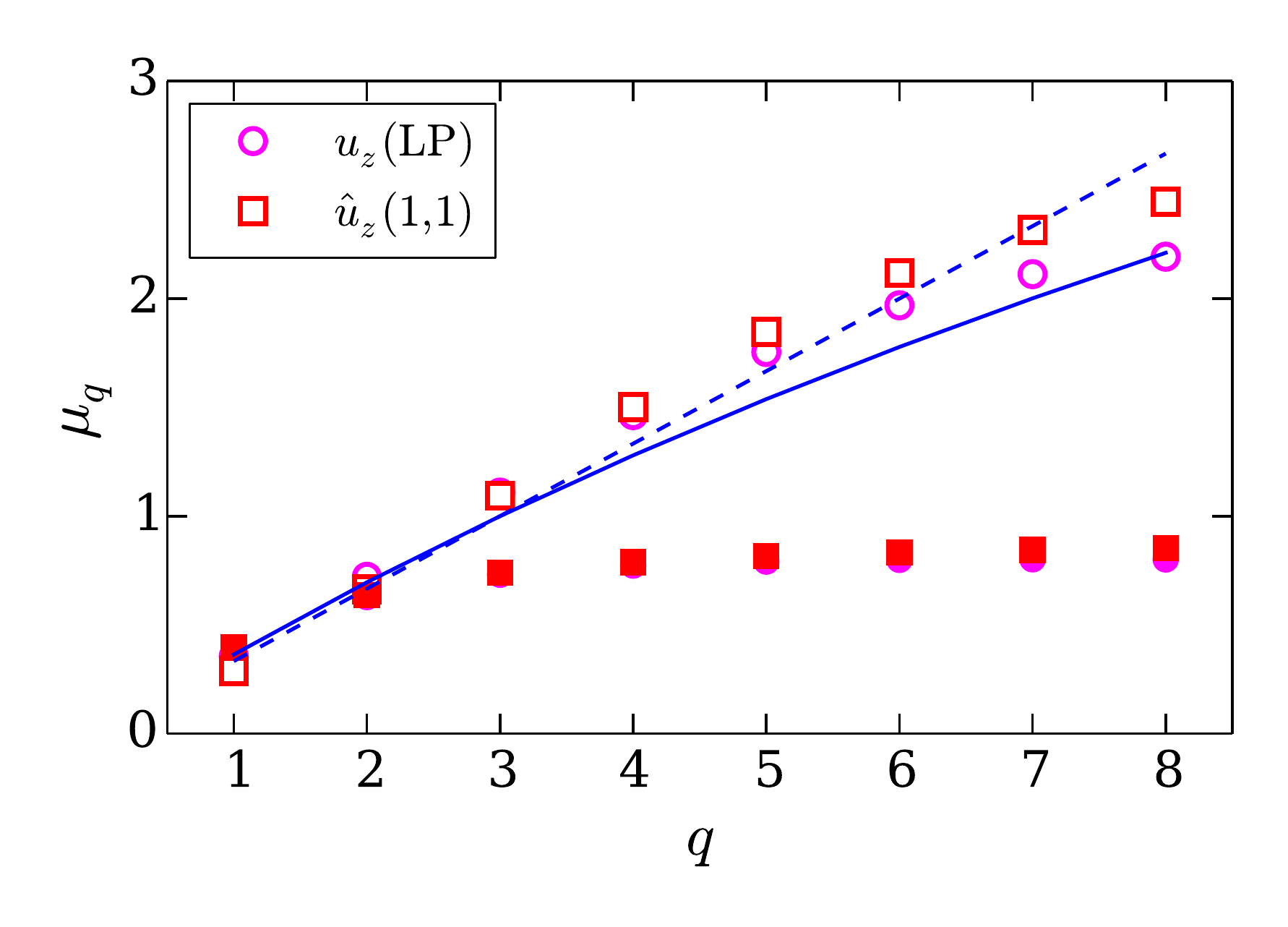}
\caption{Structure function exponents $\mu_q$ for $u_z(\mathrm{LP})$ and $\hat{u}_z(1,1)$. The $\mu_q$ increases nonlinearly with $q$ for the first scaling regime at small $\tau$  (open symbols). Blue dashed curve depicts the Kolmogorov's scaling $\mu_q = q/3$~\citep{Kolmogorov:DANS1941a}, whereas blue solid curve represents the scaling deduced by \citet{She:PRL1994}. For the second scaling regime at larger $\tau$  (filled symbols), $\mu_q$ increases markedly only up to $q = 3$, and saturates for $q \geq 4$.}
\label{fig:exp_SF}
\end{figure}

We compute the scaling exponents $\mu_q$ in the aforementioned scaling regimes, and plot them in Fig.~\ref{fig:exp_SF} as function of $q$. For the first scaling regime, we find that $\mu_q$ increases nonlinearly with increasing $q$, thus indicating that the temporal evolutions of $u_z(\mathrm{LP})$ and $\hat{u}_z(1,1)$ are intermittent. However, $\mu_q$ deviates only weakly from the linear scaling $\mu_q = q/3$, generalization of Kolmogorov's 5/3 theory~\citep{Frisch:Book} (shown as a dashed blue curve). Therefore we construe that the intermittency is not very strong. \citet{She:PRL1994} predicted a universal scaling for the spatial velocity structure functions for homogeneous and isotropic turbulence as $\mu_q = q/9 + 2[1-(2/3)^{q/3}]$ by considering a hierarchy of structures for the moments of locally averaged viscous dissipation rates. Therefore, we plot $\mu_q = q/9 + 2[1-(2/3)^{q/3}]$ as a blue solid curve in Fig.~\ref{fig:exp_SF}, and observe that our computed exponents $\mu_q$ deviate from this scaling too.
For the second scaling regime at larger delay times, $\mu_q$ saturates with increasing $q$ for $q \geq 4$ (shown as solid symbols in Fig.~\ref{fig:exp_SF}). The velocity field in  three-dimensional RBC is expected to exhibit scaling behavior similar to three-dimensional hydrodynamic turbulence~\cite{Verma:NJP2017}. It is interesting however that the temporal structure functions of the Fourier modes exhibit scaling similar to that of the velocity field. This is because the low-wavenumber Fourier modes capture the large-scale dynamics quite well.

\begin{figure}
\includegraphics[scale=0.4]{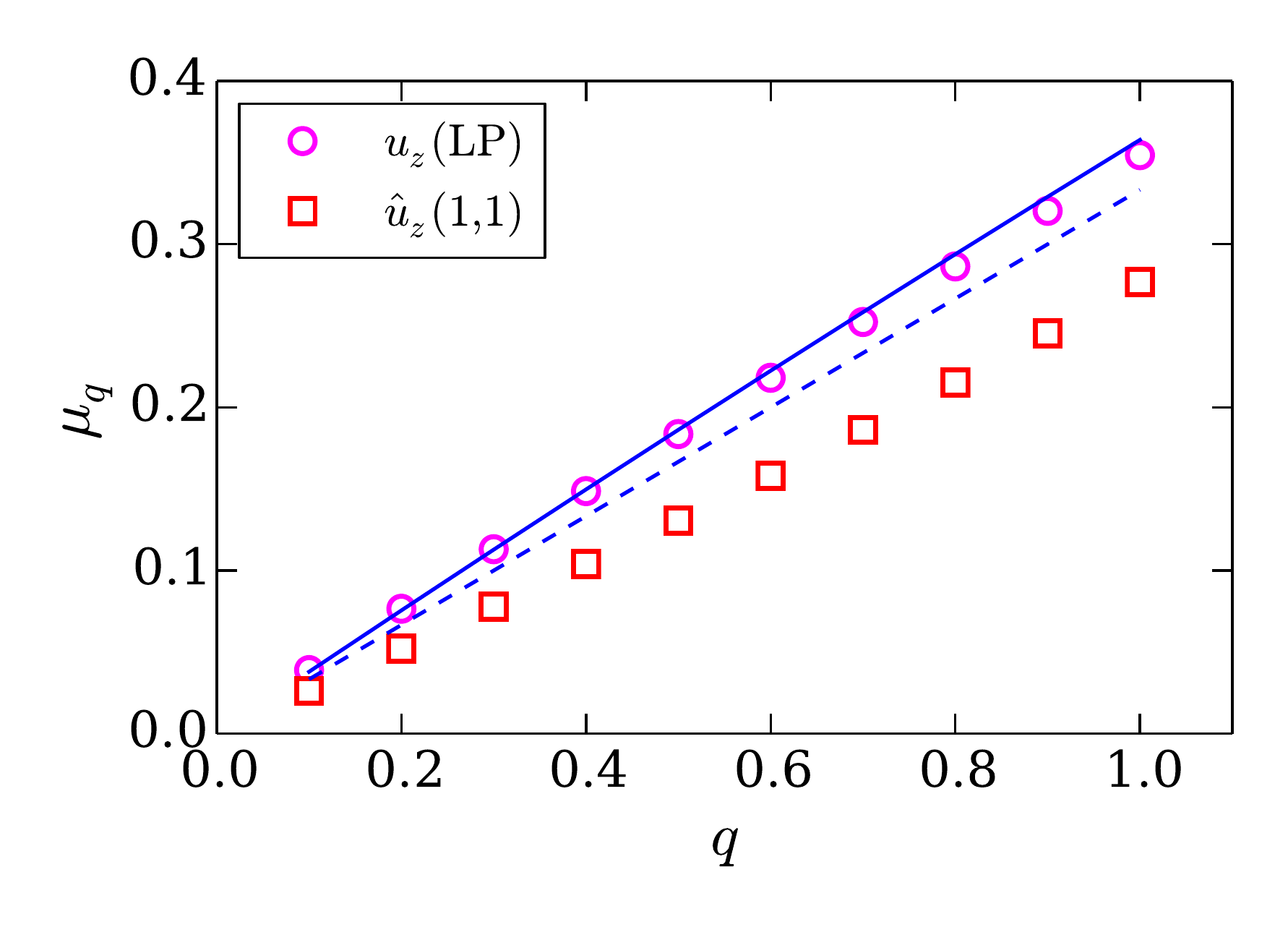}
\caption{Structure function exponents $\mu_q$ for $q \leq 1$ for $u_z(\mathrm{LP})$ and $\hat{u}_z(1,1)$. The exponents deviate from both the Kolmogorov's scaling~\citep{Kolmogorov:DANS1941a} (dashed blue curve) as well as from the She and Leveque's scaling~\citep{She:PRL1994} (solid blue curve).}
\label{fig:exp_SF_small}
\end{figure}

For a deeper understanding of the aforementioned anomalous scaling, we compute $S_q(\tau)$ for $q \leq 1$~\cite{Cao:PRL1996, Chen:JFM2005}, again identifying the aforementioned scaling regimes. The low order structure functions get contributions primarily from the core of the PDF of velocity differences, as opposed to the higher order structure functions that probe the large amplitude events present in the tails. The exponents $\mu_q$ computed for the first scaling regime are plotted in Fig.~\ref{fig:exp_SF_small}. We observe that the exponents deviate from both the Kolmogorov's scaling $q/3$ as well as from the She and Leveque's~\cite{She:PRL1994} scaling even for the low order structure functions, which is in agreement with the findings of \citet{Cao:PRL1996} and \citet{Chen:JFM2005} that the scaling is anomalous for all orders. 

\begin{figure}
\includegraphics[scale=0.35]{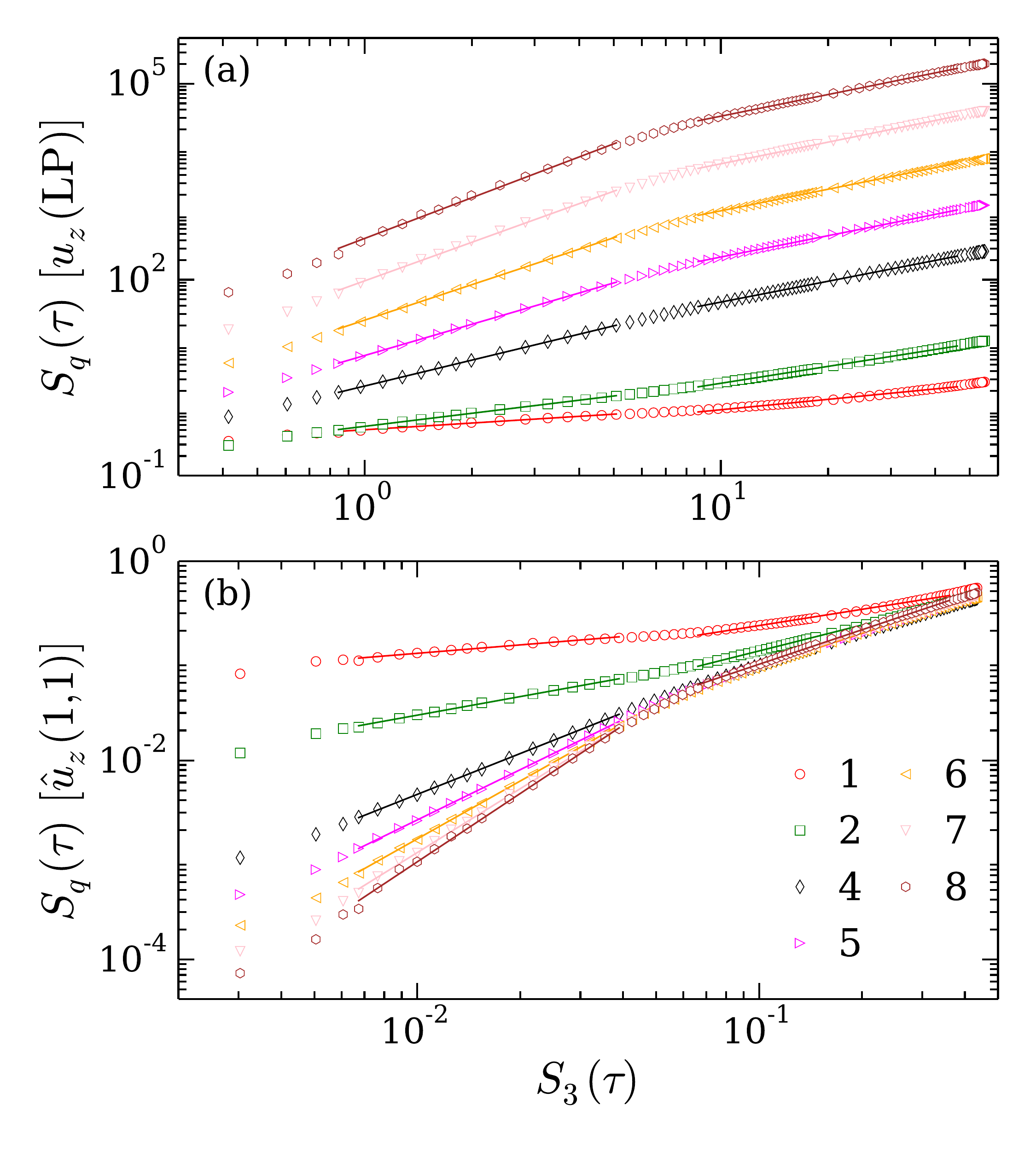}
\caption{Structure functions $S_q(\tau)$ plotted against $S_3(\tau)$ for (a) $u_z(\mathrm{LP})$ and (b) $\hat{u}_z(1,1)$. Enhanced scaling regimes compared to those in Fig.~\ref{fig:SF} can be observed here.}
\label{fig:SF_S3}
\end{figure}

\begin{figure}
\includegraphics[scale=0.32]{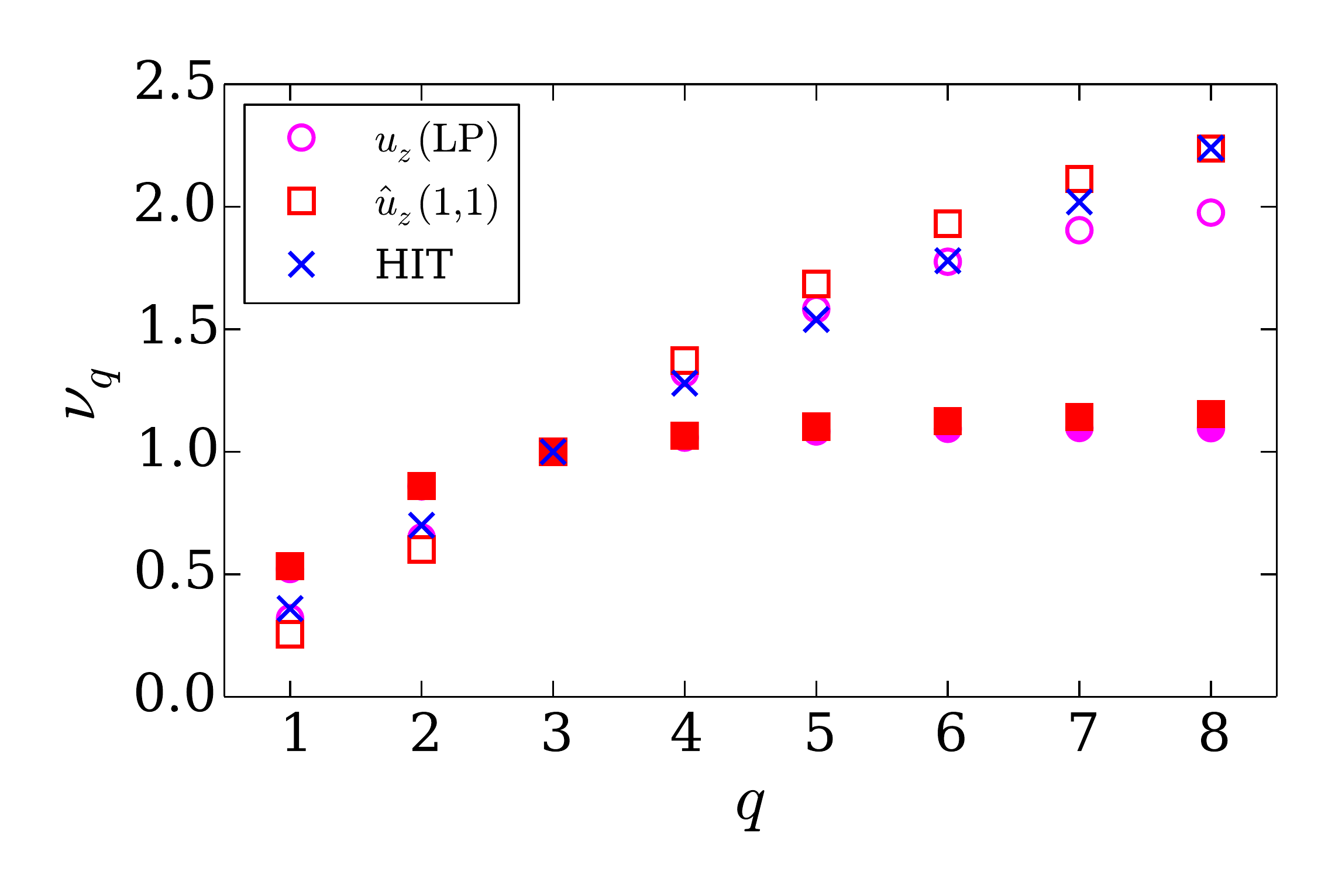}
\caption{ESS exponents $\nu_q (=\mu_q/\mu_3)$ for $u_z(\mathrm{LP})$ and $\hat{u}_z(1,1)$ for the first scaling regime at  small $\tau$ (open symbols) and for the second scaling regime at larger $\tau$ (filled symbols). Blue crosses are the ESS exponents reported in \citet{Benzi:PD1996} for the homogeneous and isotropic turbulence.}
\label{fig:exp_SF_S3}
\end{figure}

\citet{Benzi:PRE1993} proposed the extended self similarity (ESS) theory, according to which the scaling regions are enhanced if the structure functions are plotted against each other. Therefore, we plot $S_q(\tau)$ vs $S_3(\tau)$ in Fig.~\ref{fig:SF_S3} for $u_z(\mathrm{LP})$ and $\hat{u}_z(1,1)$, and find that we indeed get extended scaling regimes. The structure functions scale as $S_q(\tau) \sim [S_3(\tau)]^{\nu_q}$, with $\nu_q = \mu_q/\mu_3$. We compute $\nu_q$ for the two scaling regimes, and plot them as function of $q$ in Fig.~\ref{fig:exp_SF_S3}, which shows that, similar to $\mu_q$, $\nu_q$ also increases nonlinearly with $q$. \citet{Benzi:PD1996} reported that whereas the absolute exponents $\mu_q$ differ for different systems like RBC, magnetohydrodynamics, and homogeneous and isotropic turbulence (HIT), the relative exponents $\nu_q$ are very similar for these systems. Therefore in Fig.~\ref{fig:exp_SF_S3} we also plot $\nu_q$ for HIT reported in \citet{Benzi:PD1996} and find that they are nearly similar to $\nu_q$ determined here for $u_z(\mathrm{LP})$ and $\hat{u}_z(1,1)$. 

We would like to mention that we also computed $S_q(\tau)$ for $u_z(\mathrm{CP})$ and $\hat{u}_z(2,1)$. We however failed to detect any discernible scaling regime, specifically for higher order structure functions. For low order moments up to $q = 3$, we could recognize a very narrow regime. We also tried to use the ESS theory by plotting $S_q(\tau)$ as a function of $S_3(\tau)$, but again could not find any discernible scaling regime.

There are some similarities between phenomena of LSC under study here, and fluctuation-dominated phase ordering (FDPO)\citep{Das:PRL2000, Das:PRE2001}. The latter refers to a state which shows phase separation, in which large fluctuations lead to macroscopic rearrangements of the ordered region as a function of time. An example of a system which exhibits FDPO consists of passive particles with mutual exclusion, driven by a fluctuating surface. The behavior of long-wavelength Fourier components of the density profile in this case resemble that of the Fourier modes of the LSC system under discussion here. In both cases, the fall of the dominant long-wavelength Fourier mode in time is accompanied by a rise of the amplitude of the next few modes~\citep{Mishra:JFM2011, Chandra:PRL2013, Verma:POF2015, Das:PRL2000, Das:PRE2001, Kapri:PRE2016}, with an amplitude that decreases with increasing mode number. This signifies that both for LSC and FDPO, the system evolves within the subset of states with macroscopic structures, never reaching completely disordered states.

\begin{figure}
\includegraphics[scale=0.35]{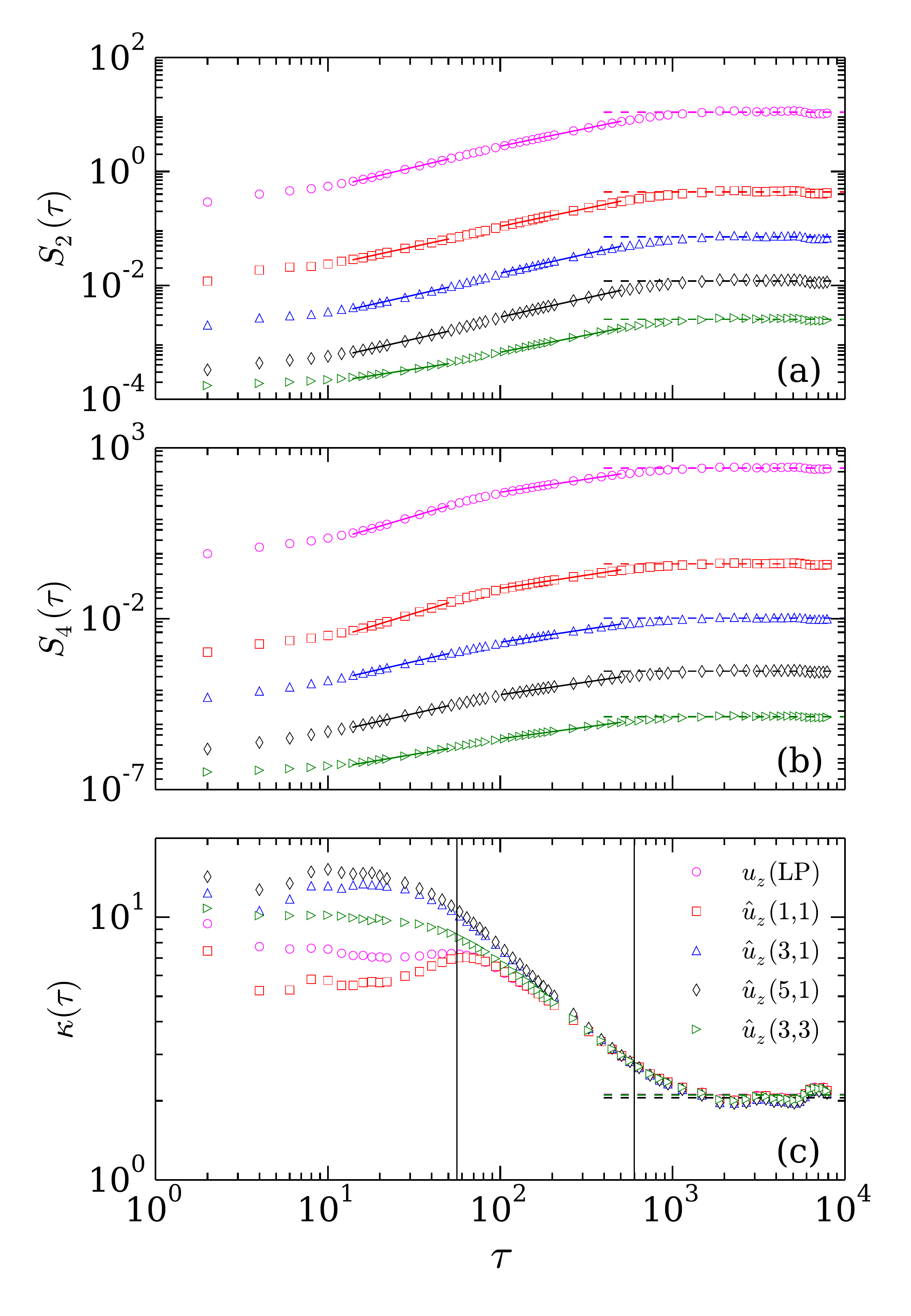}
\caption{The second order (a) and the fourth order (b) structure functions, and the flatness factor $\kappa(\tau) = S_4(\tau)/[S_2(\tau)]^2$ (c) for the vertical velocity at the left probe and for the OO modes as function of $\tau$. Dashed horizontal lines represent the estimated values of $S_2(\tau)$, $S_4(\tau)$, and $\kappa(\tau)$ in the decorrelation regime. The flatness factors are nearly constant in very large $\tau$ and small $\tau$ regimes, whereas scale as $\kappa(\tau) \sim \tau^{-0.50}$ in the intermediate $\tau$ regime indicated between two black vertical lines.}
\label{fig:S2_S4_kappa}
\end{figure}

The analogy with FDPO suggests some further directions. In their study of the passive particle system, \citet{Kapri:PRE2016} utilized the temporal structure functions to study the intermittency of the dominant Fourier mode, characterized by temporal second and fourth order structure functions. Therefore, we plot the second and fourth order structure functions in Fig.~\ref{fig:S2_S4_kappa}(a,b). It is evident that $u_z(\mathrm{LP})$ and the OO modes show similar scalings for both $S_2(\tau)$ and $S_4(\tau)$. An important quantity is the flatness factor $\kappa(\tau)$ defined as $\kappa(\tau) = S_4(\tau)/[S_2(\tau )]^2$, which is a good indicator of intermittency~\citep{Kapri:PRE2016}. In the scaling regime, $\kappa(\tau)$ scales as $\tau^{\mu_4 - 2 \mu_2}$, since we have $S_q(\tau) \sim \tau^{\mu_q}$ , and to see this we plot $\kappa(\tau)$ in Fig.~\ref{fig:S2_S4_kappa}(c). We see that the flatness factor does not vary much for the small $\tau$ regime ($\tau < 50 \, t_f$), where we observe a weak intermittency. For the intermediate regime of $\tau$ $(100 \, t_f  < \tau < 600 \, t_f)$, which corresponds to the second scaling regime, we find that $\kappa(\tau)$ varies approximately as $\tau^{-0.50 \pm 0.05}$ for $u_z(\mathrm{LP})$ and all the OO modes. This shows that the vertical velocity at the left probe and the OO modes are intermittent for the intermediate $\tau$ regime~\citep{Kapri:PRE2016, Das:PRL2000}. 

As we have discussed above, the structure functions saturate when $\tau$ is very large; this is because a signal $u(t)$ becomes uncorrelated with itself after very long time. However, the saturation values of structure functions can be predicted using the steady-state statistical properties of $u(t)$~\citep{Kapri:PRE2016}. For instance, the second order structure function in the uncorrelated regime can be estimated as $S_2(\tau) = \langle |u(t+\tau)-u(t)|^2 \rangle = 2[\langle u(t)^2 \rangle - \langle u(t) \rangle^2]$. Similarly, $S_4(\tau)$ for very large $\tau$ can be estimated as $S_4(\tau) = 2[\langle u(t)^4 \rangle - 8 \langle u(t)^3 \rangle \langle u(t) \rangle + 6 \langle u(t)^2 \rangle^2]$. We compute the values of $S_2(\tau)$ and $S_4(\tau)$ in the decorrelation regime using these relations, and indicate them as dashed horizontal lines in Fig.~\ref{fig:S2_S4_kappa}(a,b). One can observe very good agreement between the computed and predicted values.  Consequently the flatness $\kappa(\tau)$ is nearly constant in the decorrelation regime (for very large $\tau$), where we can estimate $\kappa(\tau) = 1.5 + 0.5\langle u(t)^4 \rangle/\langle u(t)^2 \rangle^2$. It is evident that a good agreement is observed between the estimated and the computed values of $\kappa(\tau)$ in the decorrelation regime. Moreover, we find that in this regime $\kappa(\tau) \approx 2$, which is less than that for a Gaussian distribution.

%%%%%%%%%%%%%%%%%%%%%%%%%%%%%%%%%%%%%%%%%%%%%%%%%%%%%%%%%%%%%%%%%%%%%%%%%%%%%%%%%%%%%%%%
%%%%%%%%%%%%%%%%%%%%%%%%%%%%%%%%%%%%%%%%%%%%%%%%%%%%%%%%%%%%%%%%%%%%%%%%%%%%%%%%%%%%%%%%

\section{Conclusions} \label{sec:conclusion}

We have studied the reversals of large-scale circulation for infinite Prandtl number RBC in a 2D square box by monitoring the vertical velocity at a probe near the left sidewall, and observed that the waiting times between two consecutive reversals are distributed exponentially~\cite{Sreenivasan:PRE2002, Brown:JFM2006, Xi:PRE2006}. Moreover, the waiting times between two consecutive ``crossings" (on shorter time scales) are distributed as a powerlaw. We observed that these exponential and powerlaw regimes are separated at $t_s \approx 40 \, t_f$. In addition, by studying the moments of generalized interswitch intervals, we observed some indication of correlation between nearby reversals, whereas there is a lack of correlation between distant reversal events. 

We also tracked the evolution of a few dominant Fourier modes of the flow, and find that the signs of all the odd-odd (OO) modes are switched after LSC reversals, while the other modes do not switch their signs. Moreover, the statistical properties of the OO modes are observed to be very similar to $u_z(\mathrm{LP})$, in particular, their integral time scales are approximately $350 \, t_f$, their probability distributions are bimodal, and their power spectra exhibit $1/f^2$ scaling for a wide range of frequencies. On the other hand, the statistical properties of $u_z(\mathrm{CP})$ are similar to those of the $\hat{u}_z(2,1)$ mode, which is because $u_z(\mathrm{CP})$ gets most dominant contribution from the $\hat{u}_z(2,1)$ mode.

Additionally, we computed the temporal structure functions for $u_z(\mathrm{LP})$ and the OO modes and found that they exhibit anomalous scaling, even for lower order structure functions. However, the intermittency is not very strong, since the scaling exponents do not deviate much from those of three-dimensional hydrodynamic turbulence~\cite{Kolmogorov:DANS1941a}. 
We also computed the flatness factor for $u_z(\mathrm{LP})$ and the OO modes, and observed that it is nearly a constant in small $\tau$ and in the decorrelation regime, whereas it scales as $\tau^{-0.50 \pm 0.05}$ in the intermediate $\tau$ regime. 

\section*{Acknowledgement}
We thank Sagar Chakraborty and Manu Mannattil for fruitful discussions. The simulations were performed on {\sc Newton} cluster at the department of Physics, IIT Kanpur.

\end{document}